\documentclass{aastex631}

\usepackage{longtable}
\begin{document}

\title{Neutrino fluxes from different classes of galactic sources}

\author{Silvia Gagliardini}
\affiliation{Department of Physics, Ariel University, Ariel, Israel}
\affiliation{Dipartimento di Fisica, Universit\`a Sapienza, P. le Aldo Moro 2, Rome, Italy}
\affiliation{Istituto Nazionale di Fisica Nucleare, Sezione di Roma, P. le Aldo Moro 2, I-00185 Rome, Italy}

\author{Aurora Langella}
\affiliation{University of Naples Federico II, Napoli, Italy and INFN Naples}
\author[0000-0002-7349-1109]{Dafne Guetta}
\affiliation{Department of Physics, Ariel University, Ariel, Israel}

\author{Antonio Capone}
\affiliation{Department of Physics, Ariel University, Ariel, Israel} 
\affiliation{Istituto Nazionale di Fisica Nucleare, Sezione di Roma, P. le Aldo Moro 2, I-00185 Rome, Italy}

\begin{abstract}
We estimate the neutrino flux from different kinds of galactic sources and compare it with the recently diffuse neutrino flux detected by IceCube. We find that the flux from these sources may contribute to $\sim 20 \%$ of the IceCube neutrino flux. 
Most of the sources selected in this work populate the southern hemisphere, therefore a detector like KM3NeT 
could help in resolving the sources out of the observed diffused galactic neutrino flux.
\end{abstract}


\section{Introduction}\label{sec:introduction}

Galactic sources are expected to accelerate cosmic rays (CRs) up to a maximum energy of $\sim 1-10$ PeV. The spectral steepening above $\sim $PeV energies results from a limit to the acceleration in the sources and from the faster escape of high energy CRs from the galaxy \citep{gaisser_2006}. Being close, these sources provide a rich environment to study the nature and mechanism of particle acceleration. Particles accelerated in galactic astrophysical sources may interact with the matter and/or with photons present in the source producing charged and neutral pions, which subsequently decay to high-energy neutrinos and 
$\gamma$-rays. Correlated production of neutrinos and 
$\gamma$-rays opens up the opportunity for multi-messenger searches. 
High energy CRs escaping the sources may interact with interstellar matter and low-energy radiation fields contained in the galactic disk producing a diffuse $\gamma$-ray emission in our Galaxy \citep{evoli,pagliaroli,lipari,cataldo,schwefer}.
Connecting high energy neutrinos to $\gamma$-ray sources is essential to reveal the presence and the amount of accelerated hadrons. In fact, since high-energy electrons can induce the production of $\gamma$-rays, via bremsstrahlung and inverse Compton scattering (ICS), the detection of $\gamma$-rays alone is not enough to distinguish between a leptonic or hadronic mechanism for the origin of the high energy photon emission. Our galaxy is densely populated with numerous high energy $\gamma$-rays point sources like Supernova Remnants (SNR), Pulsar Wind Nebulae (PWN),  Microquasars (MQ), and Novae. These sources are potential CR accelerators and therefore candidate neutrino sources.
So far, SNRs are the most promising galactic high-energy CR sources and 8 out of the 12 Galactic PeV sources detected by the  Large High Altitude Air Shower Observatory (LHAASO \citep{art:lhaaso_exp}) are compatible with SNRs \citep{lhaaso_2021}. The huge amount of kinetic energy released by a supernova, typically $10^{51}$ ergs \citep{art:dsa}, is initially carried by the expanding ejecta and is then transferred to kinetic and thermal energies of shocked interstellar gas and relativistic particles.
The mechanism of diffusive shock acceleration (DSA) can explain the ejection of relativistic particles in SNRs \citep{art:dsa}. DSA generally predicts that a substantial fraction of the shock energy is transferred to relativistic protons. 
There are evidence that cosmic-ray protons are accelerated in a SNR: \cite{art:lat_snr2} observed a characteristic pion-decay feature in the $\gamma$-ray spectra, known as the "pion-decay bump", for two SNRs (IC 443 and W44).

In the case of PWN, in order to fit the TeV emission with a purely leptonic model, the magnetic field in the nebula should be much lower than the equipartition value \citep{art:guetta_amato}.
Therefore a hadronic component is necessary to release the constrain on the magnetic field in the nebula. \cite{Palma_2017} have estimated the effects of this hadronic component to the TeV emission and the consequent neutrino flux from PWN.

The composition of microquasar jets is still an open issue, the dominant energy carrier in the jet is presently
unknown. A Doppler-shifted $H\alpha$ line has been observed in the spectrum of SS 433 that may indicate a possible presence of protons in the jets of this source.  Another diagnostic of
hadronic jets, namely the emission of TeV neutrinos, has been proposed by \cite{levinson2001}, \cite{Distefano2002}, \cite{mq_orellana} and \cite{mq_vila}.  

Novae can also be accelerators of high energy particles. High energy photons ($>100$ GeV) have been detected from the 2021 outburst of RS Oph, a recurrent nova in a symbiotic system that erupts every $\sim15$ years \citep{Acciari2022,Hess2022}.
The main interpretation of the origin of this high energy emission has been claimed to be hadronic particle acceleration in shocks \citep{Steinberg2020,Acciari2022,Hess2022}. RS Oph is the only nova detected at TeV energies so far, however there is another nova that could erupt and emit TeV emission: T Corona Borealis (T CrB)\footnote{Schaefer: private communication}. 
T CrB is amongst the all-time brightest novae, since it is amongst the closest novae to Earth (at 914 $\pm$ 23 parsecs). 
The main mechanism responsible for the high energy emission in novae is assumed to be the interaction of the nova ejecta into the pre-existing mass. TeV photons and neutrinos may be produced during this interaction \citep{Guetta_2023}.

Hadronic processes produce a roughly equal number of charged and neutral pions which decay to neutrinos and $\gamma$-rays, respectively. 
The Galactic plane has been observed in $\gamma$-rays by Fermi-LAT \citep{hunter,atwood}, by the Tibet AS$\gamma$ up to 1 PeV \citep{amenomori} and by the LHAASO experiments \citep{lhaaso_2023}. 
If part of the diffuse $\gamma$-ray emission observed in our galaxy is due to hadronic processes, a correlation with a diffuse neutrino flux emission is expected.
The energy budgets of High Energy Cosmic Rays (HECR), PeV neutrinos, and isotropic sub-TeV $\gamma$-rays are comparable \citep{fang}, supporting a strong correlation between these high-energy cosmic particles. 
The connection between $\gamma$-rays and neutrinos has motivated ANTARES to look for a possible galactic neutrino flux.
Searching for astrophysical neutrinos originated by propagation of HECR in the Galactic Ridge, ANTARES found a mild excess (in the energy range 1-100 TeV) compatible with the hypothesis that neutrinos and $\gamma$-ray fluxes have a common  origin in the same Galactic region \citep{antares2017,artsen2017,albert2018,antares2023}.

Recently, IceCube reported the observation of a diffuse neutrino emission from the Galactic plane at a 4.5$\sigma$ level of significance \citep{icecube_2023}. They found that the signal is consistent with modeled diffuse emission from the Galactic plane. Moreover, the Galactic diffuse neutrino flux agrees with multi-messenger expectations of the Galactic diffuse $\gamma$-ray flux above 400 TeV recently found by Tibet-AS and LHAASO \citep{amenomori,lhaaso_2023}.
These observations endorse the hypothesis of hadronic origin of sub-PeV diffuse $\gamma$-rays, which are generated during the propagation of tens of PeV CRs.
In the recent work of \cite{Shao:2023aoi}, the authors show that extra unresolved sources are needed to explain the LHAASO measurements in the $\gamma$-ray spectrum. They also show that a hadronic component responsible for the high energy emission cannot be excluded.

Motivated by these observations, we investigate the possibility that part of the diffuse neutrino flux may arise from a population of unresolved point sources.
We select a catalogue of galactic sources that have been detected in $\gamma$-rays at TeV energies and estimate the expected neutrino flux from each individual source from the observed high energy photon flux. We evaluate also the number of events that could be detected by IceCube and KM3NeT neutrino telescopes.

Previous studies already evaluated the expected number of neutrinos from galactic sources in IceCube (e.g. \cite{art:bednarek,art:guetta_amato,art:gonzalez,Palma_2017,art:vecchiotti_2023,art:fang_2024,art:ambrosone_2024}).
We expand these estimates by considering a larger sample of galactic sources detected at TeV energies and by predicting also the expected number of neutrino events for KM3NeT. Most of selected sources are located in the southern emisphere, where KM3NeT is expected to be more sensitive than IceCube.

In \S \ref{sec:telescopes} we describe the technical capabilities for both neutrino detectors. In \S \ref{sec:catalog} we describe and list the sources selected for our analysis.
In \S \ref{sec:nfluence} we evaluate the neutrino fluxes for each individual source. The expected number of neutrino events from each source is given in \S \ref{sec:nevents}. 
Results and conclusions are provided in \S \ref{sec:discussion}.
\newpage

\section{Neutrino Telescopes}\label{sec:telescopes}

Neutrino telescopes are based on the detection of the Cherenkov light induced by the path, in a transparent medium, of secondary relativistic charged particles produced by high energy neutrino interactions with matter. Cherenkov light is emitted at a characteristic angle, then it is possible to reconstruct both the direction and energy of the neutrinos as well as the interaction point. Neutrino detectors collect the Cherenkov light by exploiting an array of photosensors in a large volume of a medium (usually ice or water).
In the following, we report a brief description of the basic characteristics of neutrino telescopes that operate in the TeV-PeV range and that are considered in this work.\\

\noindent
\textbf{IceCube} -- The IceCube high-energy neutrino telescope is a neutrino detector located at the geographic South Pole \citep{art:detIce}. The total surface area covers roughly 1 $\rm km^2$ and it consists of 86 vertical strings arranged with 60 digital optical modules (DOMs) each, spread over depths between 1450 m and 2450 m.\\
\textbf{ANTARES} -- The ANTARES neutrino detector (2008-2020) was located in the Northern Hemisphere and was the first deep sea high energy neutrino telescope \citep{detAN}. The telescope did cover an area of less than 0.1 $\rm km^2$ anchored on the sea bed, at a depth of 2475 m, 40 km off the coast of Toulon, France. It was composed of 12 detection lines, each comprising up to 25 triplets of photo-multiplier tubes.\\ 
\textbf{KM3NeT/ARCA} -- The KM3NeT detector is the future generation deep sea Cherenkov high energy neutrino telescopes \citep{art:loi} . 
It is under construction in two different building blocks, each one consisting of 115 strings equipped with 18 DOMs. 
KM3NeT/ARCA  will be anchored at a depth of 3500 m at a site 80 km South-East of Portopalo di Capo Passero, Sicily, Italy.
KM3NeT/ARCA will have large spacings between adjacent strings about 100 m in order to target astrophysical neutrinos at TeV energies. 
The detector is already in operation with about 30 strings while it is under construction.

Each detector can reconstruct the direction of high energy particles originated from the neutrino interactions. Charged Current neutrino events provide high energy ($\geq$ 1 TeV) muons in the final state that, in water or ice, can travel for kms. The direction of muon is a good proxy of the incoming neutrino direction and indicates the position of the astrophysical neutrino source. Neutral current neutrino events originate showers that, if contained in the detector, allow good reconstruction of the event energy, but an approximate reconstruction of the incoming neutrino direction. 
In this work we base our evaluations on the estimate of track-like events in order to have the best angular resolution and a better background rejection.

\begin{figure}[h!]
    \centering
    \includegraphics[scale=0.5]{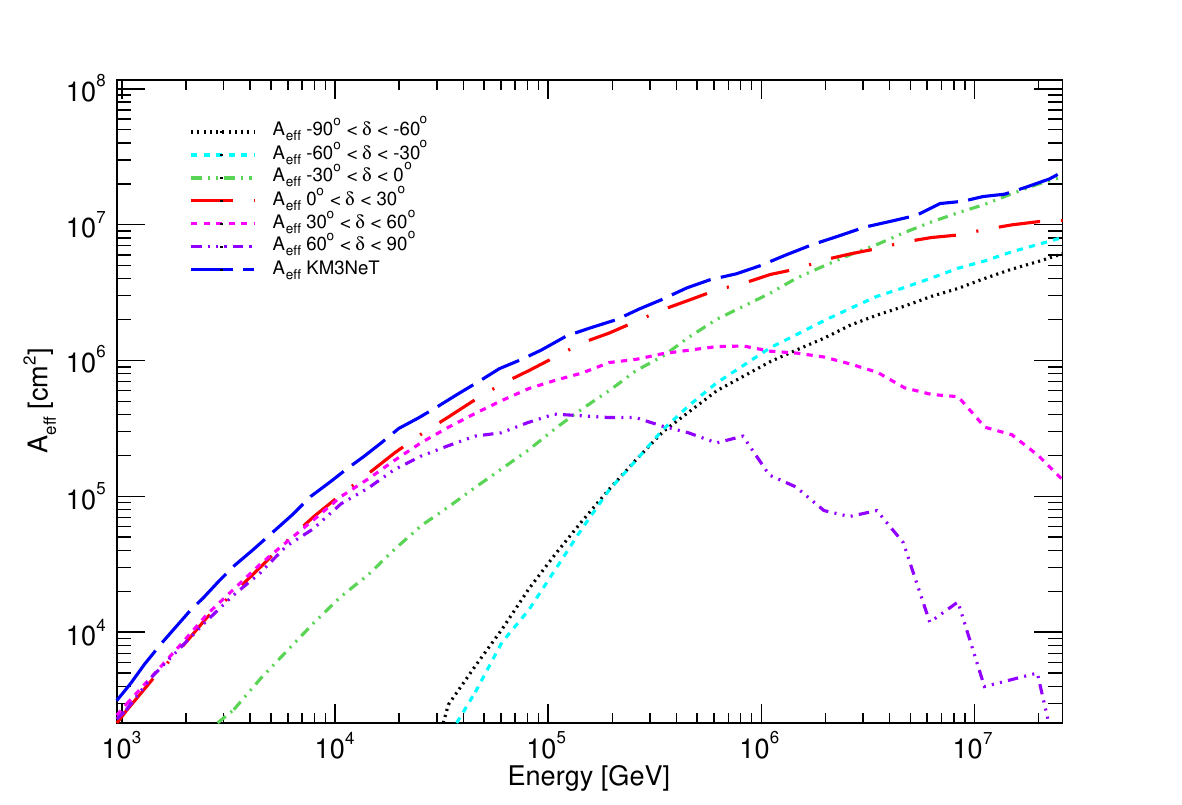}
    \caption{Effective areas for track events ($\nu_{\mu}$) of Icecube \citep{art:icecube_aeff} and KM3NeT \citep{art:loi} as a function of the energy. For Icecube the effective area is given as a function of the declination.}
    \label{fig:effective_areas}
\end{figure}

In Fig.\ref{fig:effective_areas} we show IceCube and KM3NeT effective areas for neutrino induced events collected as tracks 
in the high energy region \citep{art:icecube_aeff,art:loi}. The effective areas are given as a function of the energy and for IceCube also as a function of the declination. For KM3NeT the source declination is taken into account as the fraction of time in which it is visible below the horizon \citep{art:km3_visibility}.

\section{Description of the selected galactic sources sample}\label{sec:catalog}

For the present study we select, from the TeVCat catalogue \citep{tev_cat}, galactic sources with these characteristic: (i) the spectrum has been measured in the TeV energy range and (ii) it can be described by a Power Law (PL) as:
\begin{equation}
    \frac{dN_{\gamma}}{dE_{\gamma}}=N_0 \left(\frac{E}{E_0}\right)^{-\alpha}
    \label{eq:pl}
\end{equation}

where $N_{0}$ is the differential flux at a reference (pivot) energy $E_0$ in TeV$^{-1}$ cm$^{-2}$ s$^{-1}$ and $\alpha$ the spectral index for each source. $N_0$, $E_0$ and $\alpha$ are taken from the catalogue.\\
Our sample includes different kind of sources: PWN, SNR, Molecular Cloud, TeV Halo, Binary system, Microquasar, Novae, Superbubble and Massive Stellar Cluster as well as some unidentified sources. 

PWNe are the result of the explosion of a supernova characterized by a diffuse nebulae powered by the wind of a fast-spinning, highly magnetised neutron star. The Crab Nebula is the youngest and most energetic PWN known and it is considered as a reference source for PWN studies.

SNRs are the result of the explosion of a supernova: an explosion which drives its progenitor material supersonically into interstellar space, forming a collisionless shock wave ahead of the stellar ejecta. SNRs are characterized by a shell-like structure resulting from the explosion of a star in a supernova event.  In this case the emission comes from the shocked material that has a shell-type (Shell) structure. 
We consider also composite SNR (Comp. SNR) that are supernova remnant of a binary star system where there is a combination of a white dwarf and a massive companion star.
Some SNRs originate from core-collapse supernova explosions of massive stars in molecular clouds (MC). 
They are associated with the relative MC \citep{SNR_MC}. 
For this reason, some sources in our sample are listed as SNR/MC. 

TeV Halos (here abbreviated as TH) are a relatively new class of sources, firstly introduced to described Milagro's observation of an extended TeV $\gamma$-ray emission surrounding the nearby Geminga pulsar, confirmed also by the High Altitude Water Cherenkov (HAWC) observatory \citep{Geminga, tevhalo_hawc, art:HAWC_cat}. TeV Halos are found to be coincident in position with pulsars or PWNe. However, since these sources appear morphologically and dynamically distinct from PWNe detected in X-ray and radio observations, they have been classified by  \cite{linden} as a new $\gamma$-ray source class.

Some binaries are also included in the catalogue as well as Microquasars (MQ), i.e. X-ray binaries characterized by non-thermal jets. These systems are believed to consist of a compact object, a neutron star or a black hole, and a companion star. The name Microquasar derived from the similarity in terms of the physical processes with extragalactic Quasars, even though the scale where these processes happen is smaller.

Novae are powerful eruptions following a thermonuclear runaway (TNR) that occurs below the surface of a White Dwarf (WD) \citep{Starrfield1972,Shara1981,Starrfield2008}. During its cycle (accretion, eruption and decline) a Nova producing system could possibly be observed in the infrared (IR), ultraviolet (UV), soft and hard X-rays and even $\gamma$-rays.
We consider two Novae in our sample, RS Oph that was detected at TeV energy \citep{Hess2022} and T CrB. The latter has not yet been detected at TeV energies, then we predict its neutrino flux assuming that its TeV $\gamma$ spectrum will be similar to the one of RS Oph. T CrB is three time closer than its sister-nova RS Oph, so the T CrB neutrino flux is expected to be nine times much brighter. T CrB is in the northern sky at declination +25$^0$ therefore is an ideal candidate for the IceCube telescope but visible also for KM3NeT.

Another TeV source considered in our sample is ARGO J2031+4157, identified as Superbubble, i.e. a cavity populated with hot gas atoms, probably due to multiples Supernovae and stellar winds. In particular, according to \cite{argo}, it can be considered \textbf{the TeV counterpart} of the  Cygnus Cocoon.
 
Finally, three sources (HESS J1848-018, Westerlund 1 and Westerlund 2) are Massive Stellar Cluster (here abbreviated as MSC) that have been detected at TeV \citep{art:hpgs2018}.
Young massive clusters are dense aggregates of young stars that form the fundamental building blocks of galaxies. Given the high energy photons detected by these sources, they have been considered as potential accelerator of high energy particle \citep{Mohrmann2021}.

In Table \ref{long1} we list all the selected Galactic sources that have been detected at TeV energies.

  \begin{longtable}[c]{| c | c | c | c | c | c | c | c |}
  \caption{Spectral properties of the selected catalogue. A Power Law spectrum (Eq.\ref{eq:pl}) is used for each source according to the parameters listed in the catalogue. Sources included are PWN/TeVHalo (PWN/TH), Shell-like SNR(Shell), SNR/MolecularCloud (SNR/MC), Composite SNR (comp.SNR),   Superbubble (SB), Massive Star Cluster (MSC) as well as some unidentified sources (unid). \label{long1}}
  \\
 \hline
 \multicolumn{8}{| c |}{Begin of Table}\\
 \hline
  \textit{Source} & \textit{Type} & \textit{$\delta$} & $N_0$ & $\alpha$ & $E_0$  & \textit{Distance} & \textit{Ref.}\\

  \textit{Name}&  & [degree] & [$TeV^{-1} cm^{-2} s^{-1}$] & & [$TeV$] & [$kpc$] & \\
 \hline
 \endfirsthead

 \hline
 \multicolumn{8}{|c|}{Continuation of Table 1}\\
 \hline
  \textit{Source} & Type & declination & $N_0$ & $\alpha$ & $E_0$  & \textit{Distance} & \textit{Ref.}\\
  \textit{Name}&  & [degree] & [$TeV^{-1} cm^{-2} s^{-1}$] & & [$TeV$] & [$kpc$] & \\
 \hline
 \endhead

 \hline
 \endfoot

 \hline
 \multicolumn{8}{| c |}{End of Table}
 \\
 \hline
 
 \hline
 \endlastfoot

Geminga  &   PWN/TH  &   17.76   &   $3.77\cdot10^{-16}$   &   2.6   &   35    &   0.3  & \citep{Geminga}\\ 
Vela X  &   PWN/TH  &  -45.66  &   $1.83\cdot10^{-12}$   &   1.89   &   3.02    &   0.3 &\citep{art:hpgs2018} \\ 
Boomerang  &   PWN  &  61.24  &   $7.09\cdot10^{-16}$   &   2.6   &   35    &   0.8 & \citep{Geminga} \\ 
SNR G106.3+02.7  &   Shell  &   60.88  &   $1.15\cdot10^{-13}$   &   2.29   &   3    &   0.8 & \citep{art:SNRG106.3} \\ 
T-CrB & Nova & 25.92 & $4.75 \cdot 10^{-11}$ & 3.33 & 0.35 & 0.9 & * \\
ARGO J2031+4157  &   SB &   42.5   &   $2.5\cdot10^{-11}$   &   2.6   &   1    &   1.4  & \citep{argo}\\ 
CTA 1  &   PWN  &   72.98   &   $9.1\cdot10^{-11}$   &   2.2   &   3    &   1.4 & \citep{CTA1}  \\ 
HESS J0632+057  &   Binary  &   5.79   &   $4.56\cdot10^{-12}$   &   2.67   &   0.5    &   1.4 & \citep{HESSJ0632}  \\ 
IC 443  &   Shell  &   22.53   &   $1.00\cdot10^{-11}$   &   3.1   &   0.4    &   1.5 &\citep{IC443}  \\ 
TeV J2032+4130  &   PWN  &   41.58   &   $2.1\cdot10^{-14}$   &   3.22   &   10    &   1.8  & \citep{TeVJ2032+4130}\\ 
PSR J2032+4127 &   Binary  &   41.46   &   1.45$\cdot10^{-12}$   &   2.73   &   1    &   1.8  & \citep{PSRJ2032+4127}\\ 
LS I +61 303  &   MQ  &   61.27  &   $4.4\cdot10^{-13}$   &   2.4   &   1    &   2.0 & \citep{art:LS+61_tev}  \\ 
Crab  &   PWN  &   22.01   &   $2.83\cdot10^{-11}$   &   2.62   &   1    &   2.0 & \citep{Crab} \\ 
3C 58  &   PWN  &   64.85   &   $2\cdot10^{-13}$   &   2.4   &   1    &   2.0 &\citep{3c58} \\  
HESS J1800-240A  &   SNR/MC  &   -24.02  &   $4.8\cdot10^{-12}$   &   2.47   &   0.95    &   2.0  &\citep{art:hpgs2018} \\ 
W28  &   SNR/MC  & -23.34 &   $7.5\cdot10^{-13}$   &   2.66   &   1.00    &   2.0  & \citep{art:hpgs2018}\\ 
SN 1006 NE  &   Shell  &   -41.8   &   $4.65\cdot10^{-13}$   &   2.35   &   1    &   2.2  & \citep{art:SN1006}\\ 
Cygnus X-1  &   MQ  &   35.20   &   $2.3\cdot10^{-12}$   &   3.2   &   1    &   2.2 &\citep{art:cygx1_tev} \\ 
HESS J1026-582  &   PWN  & 	-58.23  &   $3.64\cdot10^{-13}$   &   1.81   &   4.43    &   2.3  &\citep{art:hpgs2018}\\ 
HESS J1708-443  &   PWN  & 	-44.29 &   $1.23\cdot10^{-12}$   &   2.17   &   1.70    &   2.3  & \citep{art:hpgs2018}\\ 
HESS J1356-645  &   PWN  & -64.51 &   $5.73\cdot10^{-13}$   &   2.2   &   2.74    &   2.4  & \citep{art:hpgs2018}\\ 
LS 5039  &   MQ  &  -14.85  &   $9.82\cdot10^{-13}$   &   2.32   &   1.05    &   2.5 & \citep{art:hpgs2018} \\ 
PSR B1259-63  &   Binary  & -63.85 &   $2.62\cdot10^{-13}$   &   2.59   &   1.40    &   2.7 & \citep{art:hpgs2018}  \\ 
HESS J1729-345  &   unid &   -34.53  &   $8.25\cdot10^{-13}$   &   2.43   &   1.16    &   3.2  & \citep{art:hpgs2018}\\ 
Westerlund 1  &   MSC  & -46.26 &   $7.87\cdot10^{-12}$   &   2.54   &   1.05    &   3.2 & \citep{art:hpgs2018}   \\ 
HESS J1731-347  &   Shell  &  -34.76 &   $4.67\cdot10^{-12}$   &   2.32   &   0.78    &   3.2 & \citep{art:hpgs2018} \\ 
SNR G318.2+00.1  &   SNR/MC  & 	-59.46 &   $2.2\cdot10^{-12}$   &   2.52   &   1.54    &   3.5  & \citep{art:hpgs2018}\\ 
Tycho  &   Shell  &   64.14  &   $2.2\cdot10^{-13}$   &   2.92   &   1    &   3.5 &\citep{Tycho} \\ 
HESS J1809-193  &   unid  &   -19.33   &   $6.62\cdot10^{-12}$   &   2.38  &   1.05    &   3.7  & \citep{art:hpgs2018}\\ 
HESS J1825-137  &   PWN/TH  &   -13.97   &   $1.72\cdot10^{-11}$   &   2.38   &   1.16    &   3.9  & \citep{art:hpgs2018}\\ 
HESS J1826-130  &   unid  &   -13.02   &   $2.73\cdot10^{-13}$   &   2.04   &   2.06    &   4.0  & \citep{art:hpgs2018}\\ 
HESS J1834-087  &   unid  &  -8.75   &   $5.79\cdot10^{-12}$   &   2.61   &   0.87    &   4.0 & \citep{art:hpgs2018} \\ 
HESS J1718-385  &   PWN  &  -38.51   &   $4.01\cdot10^{-14}$   &   1.78   &   4.02    &   4.2  & \citep{art:hpgs2018}\\ 
RS Ophiuchi & Nova & -6.7 & $2.38 \cdot 10^{-12}$ & 3.33 & 0.35 & 4.2 & \citep{Hess2022}\\
HESS J1813-178  &   PWN  &   -17.83   &   $1.01\cdot10^{-12}$   &   2.07   &   1.40    &   4.7 & \citep{art:hpgs2018} \\ 
SNR G015.4+00.1  &   Comp.SNR  &  -15.47   &   $1.07\cdot10^{-13}$   &   2.21   &   1.54    &   4.8  & \citep{art:hpgs2018}\\ 
HESS J1833-105  &   PWN  & 	-10.57   &   $4.22\cdot10^{-13}$   &   2.42   &   0.95    &   4.8  & \citep{art:hpgs2018} \\ 
SNR G292.2-00.5  &   PWN  &  -61.45  &   $7.96\cdot10^{-13}$   &   2.64   &   1.27    &   5.0  & \citep{art:hpgs2018} \\ 
MSH 15-52  &   PWN  & -59.16   &   $2.58\cdot10^{-12}$   &   2.26   &   1.54    &   5.2  & \citep{art:hpgs2018} \\ 
HESS J1848-018  &   MSC  & 	-1.89  &   $2.51\cdot10^{-12}$   &   2.57   &   0.87    &   5.3  & \citep{art:hpgs2018}\\ 
W 51  &   SNR/MC  &  14.14   &   $4.52\cdot10^{-13}$   &   2.55   &   1.40    &   5.4  & \citep{art:hpgs2018}\\ 
SS 433 w1 & MQ & 5.04 & $8.6\cdot10^{-14}$ & 2.54 & 3 &  5.5 & \citep{lhaaso_cat}\\ 
SS 433 e1 & MQ & 4.93 & $4.5\cdot10^{-17}$ & 3.3 & 50 & 5.5 & \citep{lhaaso_cat} \\ 
Kookaburra (PWN)  &   PWN  &  -60.75  &   $8.41\cdot10^{-13}$   &   2.20   &   1.87    &   5.9  & \citep{art:hpgs2018} \\ 
HESS J1804-216  &   unid  &  -21.74  &   $2.13\cdot10^{-11}$   &   2.69   &   0.72    &   6.2  & \citep{art:hpgs2018}\\ 
SNR G054.1+00.3  &   PWN  & 18.84 &   $1.28\cdot10^{-13}$   &   2.59   &   1.70    &   6.2  & \citep{art:hpgs2018}\\ 
HESS J1846-029  &   PWN  & 	-2.97  &   $5.96\cdot10^{-13}$   &   2.41   &   1.05    &   6.3  & \citep{art:hpgs2018}\\ 
HESS J1616-508  &   PWN  & 	-50.86  &   $7.55\cdot10^{-12}$   &   2.32   &   1.16    &   6.5  & \citep{art:hpgs2018}\\ 
HESS J1303-631  &   PWN  & 	-63.19  &   $1.53\cdot10^{-12}$   &   2.33   &   1.87    &   6.6  & \citep{art:hpgs2018}\\ 
HESS J1837-069  &   PWN  &  -6.96 &   $2.00\cdot10^{-11}$   &   2.54   &   0.95    &   6.6  & \citep{art:hpgs2018}\\ 
IGR J18490-0000  &   PWN  &  -0.04  &   $7.66\cdot10^{-14}$   &   1.97   &   2.74    &   7.0  & \citep{art:hpgs2018}\\ 
CTB 37A  &   SNR/MC  & -38.52  &   $2.76\cdot10^{-13}$   &   2.52   &   1.05    &   7.9  & \citep{art:hpgs2018}\\ 
Westerlund 2  &   MSC  & -57.78 &   $7.50\cdot10^{-13}$   &   2.36   &   1.87    &   8.0  & \citep{art:hpgs2018}\\ 
SNR G000.9+00.1  &   PWN  &  -28.15   &   $8.38\cdot10^{-13}$   &   2.40   &   1.00    &   8.5  & \citep{art:hpgs2018}\\ 
HESS J1640-465  &   Comp.SNR  & -46.57  &   $4.59\cdot10^{-12}$   &   2.57   &   0.95    &   8.6  & \citep{art:hpgs2018}\\ 
HESS J1634-472  &   unid  & -47.23  &   $1.73\cdot10^{-12}$   &   2.31   &   1.40    &   8.6  & \citep{art:hpgs2018}\\ 
SNR G327.1-01.1  &   PWN  & -55.09  &   $5.79\cdot10^{-14}$   &   2.19   &   2.26    &   9.0 & \citep{art:hpgs2018} \\ 
W 49B  &   SNR/MC  &  9.09   &   $3.15\cdot10^{-13}$   &   3.14   &   1.00    &   11.3 & \citep{art:hpgs2018} \\ 
SNR G349.7+00.2  &   SNR/MC  &  -37.44  &   $2.25\cdot10^{-13}$   &   2.80   &   1.00    &   11.5  & \citep{art:hpgs2018}\\ 
CTB 37B  &   Shell  &  	-38.22 &   $1.74\cdot10^{-12}$   &   2.74   &   0.79    &   13.2 & \citep{art:hpgs2018} \\

 \end{longtable}




\section{Galactic diffuse neutrino flux estimate}\label{sec:nfluence}

In this section, we derive the maximum  neutrino flux expected from the galactic sources considered in this paper. In order to obtain the maximum neutrino flux, we assume that the entire high energy photon emission detected from these sources at $\sim$TeV energies is due only to hadronic processes and we convert the TeV flux to a neutrino flux. 

Relativistic protons may produce high energy $\gamma$-rays (1-100 TeV) either by inelastic nuclear collisions or by photo-meson interactions. The dominant mechanism depends on the relative density between target photons and protons present inside the source. 
Processes similar to the ones (p-p collision or photomeson interaction) that produce the neutral pions ($\pi^0$), and subsequently the sub-TeV photons, would also generate charged pions ($\pi^\pm$) that decay into high energy neutrinos:
\begin{equation}
p+p \rightarrow \pi^0, \pi^+, \pi^- p, n, ...
\end{equation} 

\begin{equation}
p + \gamma \rightarrow \Delta^+ \rightarrow  p+\pi^0,  n+\pi^+ 
\end{equation} 

These processes, taking into account also the subdominant ones \citep{povh_2004}, give in the final state almost the same number of charged and neutral pions. Neutral pions decay into two $\gamma$-rays: 

\begin{equation}
\pi^0\rightarrow \gamma + \gamma
\end{equation}

while for charged pions decays we get:

\begin{equation}
\pi^+ \rightarrow \mu^+ + \nu_{\mu} \rightarrow e^+ + \nu_e + \bar{\nu}_{\mu}+ \nu_{\mu}
\end{equation}

\begin{equation}
\pi^- \rightarrow \mu^- + \bar{\nu}_{\mu} \rightarrow e^- + \bar{\nu}_e + \nu_{\mu} + \bar{\nu}_{\mu}
\end{equation}
where $\nu_\mu$ and $\nu_e$ are the muon and electron neutrinos respectively.
Averaging over the neutrino flavours we obtain an approximate relation between the photon and the neutrino fluxes similar to what is given in \cite{razzaque_2010,Yacobi_2014,Palma_2017}:

\textbf{\begin{equation}\label{eq:pions}
\frac{dN_{\nu+\bar{\nu}}}{dE_{\nu+ \bar{\nu}}} \approx \frac{dN_{\gamma}}{dE_{\gamma}}
\label{eq:flussi}
\end{equation}
}

and therefore
\begin{equation}
\int_{E_{\nu + \bar{\nu}}^{\rm min}}^{E_{\nu + \bar{\nu}}^{\rm max}}E_{\nu + \bar{\nu}}\frac{dN_{\nu + \bar{\nu}}}{dE_{\nu + \bar{\nu}}}dE_{\nu + \bar{\nu}}= \int_{E_{\gamma}^{\rm min}}^{E_{\gamma}^{\rm max}}E_{\gamma}\frac{dN_{\gamma}}{dE_{\gamma}}dE_{\gamma}
\end{equation}
where $ E_{\gamma}^{\rm min}$=$E_{\nu + \bar{\nu}}^{\rm min}$ and $E_{\gamma}^{\rm max}$=$E_{\nu + \bar{\nu}}^{\rm max}$ are the minimum and maximum photon/neutrino energies respectively. From now on with "$\nu$" we will refer to the sum of neutrinos and antineutrinos.
For the sources selected in Table \ref{long1}, the high energy photon emission is represented by a single Power Law (Eq.\ref{eq:pl}), then following Eq.\ref{eq:flussi} we can assume the same PL also for neutrino flux.


For each neutrino energy $E_k$ we evaluate the flux from a single source \textit{i} in the following way:

\begin{equation}
(d\Phi_k)_{i}= N_{0,i} (\frac{E_k}{E_{0,i}})^{-\alpha_{i}}    
\end{equation}

Then, for $E_{\nu}= E_k$, we sum the neutrino fluxes over all the sources considered in the sample defined in Table \ref{long1} and we compare the result with the galactic neutrino flux observed by IceCube in \cite{icecube_2023}:

\begin{equation}
    d \Phi_{k}=  \sum_{i=1}^{M} N_{0,i} \big(\frac{E_k}{1 TeV}\big)^{- \alpha_{i}} 
\end{equation}

where $M$ is the total number of sources, $N_{0,i}$ the normalization coefficient of the \textit{i} source in TeV$^{-1}$ cm$^{-2}$ s$^{-1}$ and $\alpha_i$ the spectral coefficient for each source.

\begin{figure}
    \centering
    \includegraphics[scale=0.5]{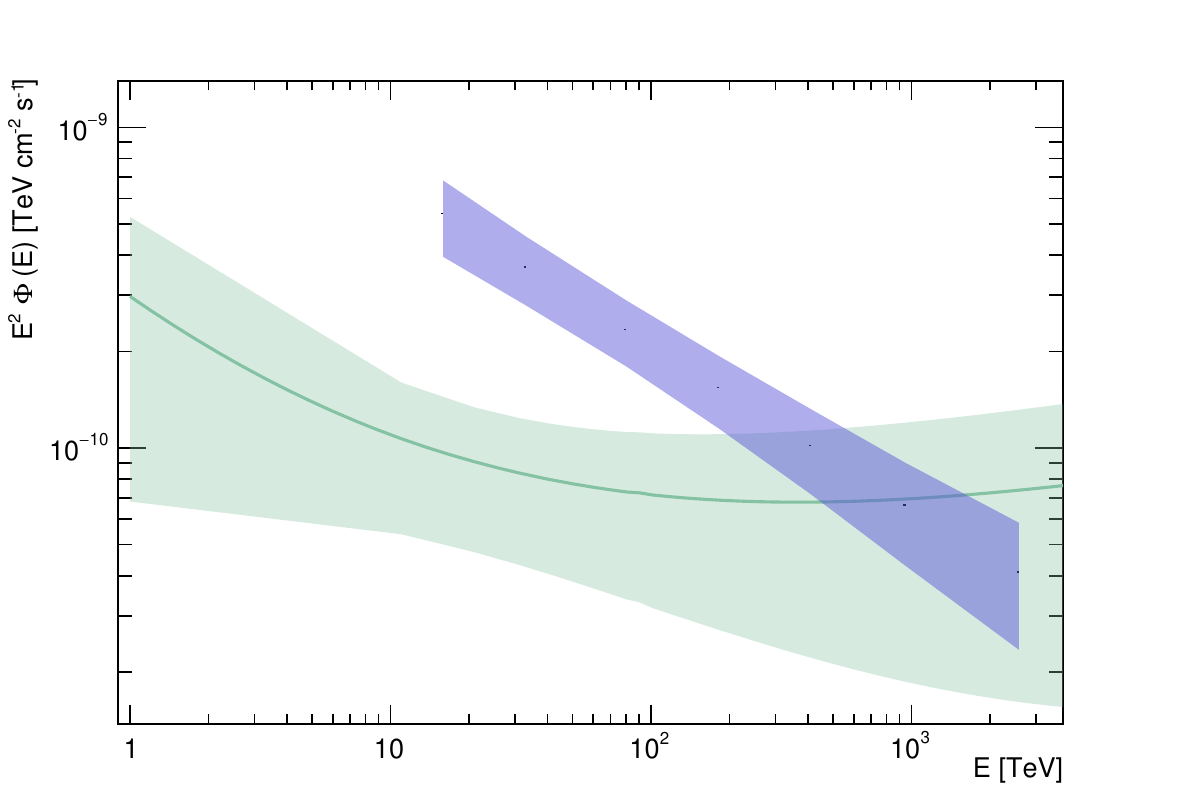}
    \caption{Total neutrino spectral energy distribution (SED) due to galactic point sources in green, compared with the IceCube Galactic diffuse neutrino flux \citep{icecube_2023} in blue. 
    }
    \label{fig:sed_sum}
\end{figure}

In Figure \ref{fig:sed_sum} we show the obtained Spectral Energy Density (SED), relative to the neutrino total flux, and we compare the obtained result with the IceCube Galactic diffuse neutrino flux as reported in \cite{icecube_2023}. The error band associated to our calculation and shown in Figure \ref{fig:sed_sum} is obtained considering the uncertainties on the parameters $\alpha$ and $N_0$ for each single source.



By looking at the plot we see that the galactic flux estimated in this work is $\sim 20\%$ of the IceCube All-Sky $\nu$ Flux for $E\sim 10$ TeV. 

\section{Neutrino fluxes from point like galactic sources}\label{sec:nevents}

We can now compute the high energy neutrino flux at Earth from each point like source listed in Table \ref{long1} and estimate the number of events that may be detected by the telescopes described in \S\ref{sec:telescopes}:

\begin{equation}\label{eq:integral}
n_{events}=\int_{E_1}^{E_2} T \frac{dN_{\nu}}{dE_{\nu}}A(E_{\nu})dE_{\nu} 
\end{equation}

where $T$ is the exposure time of a neutrino telescope, $dN_{\nu}/dE_{\nu}$ can be derived from $dN_{\gamma}/dE_{\gamma}$ according to Eq.\ref{eq:pions}, $A(E_{\nu})$ is the effective area of the considered neutrino telescope and [$E_1 \div E_2$] is the energy range [$1 \div 1000$] TeV. 
The number of neutrino interactions expected in IceCube, for which the effective area is known \citep{art:icecube_aeff} as a function of the energy and of the source declination $A(E_{\nu},\delta)$, is estimated as:
\begin{equation}\label{eq:integral2}
    n_{events}^{IC}= \int_{E_1}^{E_2} T \frac{dN_{\nu}}{dE_{\nu}} A(E_{\nu}, \delta) dE_{\nu}  
\end{equation}

The number of neutrino interactions expected in KM3NeT/ARCA full detector, made by two building blocks, is obtained  knowing the total effective area \citep{art:loi} and taking into account the source visibility $V(\delta)$ characterized by its declination $\delta$ as given by \cite{art:km3_visibility}:
\begin{equation}\label{eq:integral3}
    n_{events}^{ARCA}= \int_{E_1}^{E_2} T \frac{dN_{\nu}}{dE_{\nu}} A(E_{\nu}) V(\delta) dE_{\nu}  
\end{equation}

The main background component is due to the flux of atmospheric neutrinos resulting from the interaction of cosmic rays, high-energy protons and nuclei, with the atmosphere. Atmospheric neutrinos and muons are the products of the decays of charged mesons originated in atmosphere by cosmic-ray interactions. 
The atmospheric neutrino flux above 1 TeV is expressed as a power law:

\begin{equation}
    \frac{d \Phi_{\nu_{atm}}}{dE_{\nu_{atm}} d \Omega}= C_{\nu_{atm}} E_{\nu_{atm}}^{-\beta}
\end{equation}

We obtain the parameter $C_{\nu_{atm}}$ and $\beta$ by fitting the Zenith-averaged, unfolded atmospheric muon neutrino energy spectrum from 100 GeV to 400 TeV \citep{art:atmosferici}:

\begin{equation}
    \Phi_{\nu_{\mu}, atm}(E_{\nu})= 2.64 \cdot 10^{-8} E_{\nu}^{-3.4} cm^{-2} s^{-1} sr^{-1} TeV^{-1}
    \label{eq:atmosferici}
\end{equation}

The background due to atmospheric $\nu$ interactions cannot be identified on an event by event basis. Its contribution to the data sample can be evaluated and subtracted on a statistical base. 
For each detector (Icecube and KM3NeT/ARCA), we evaluate the number of background events that will be collected in the same search cone opened around the source under investigation:

\begin{equation}
   n_{BG}^{IC}= \int_{\Delta \Omega} \int_{E_1}^{E_2} 2 T \frac{d \Phi_{\nu}}{dE_{\nu} d \Omega} A(E_{\nu}, \delta) dE_{\nu} d \Omega 
\end{equation}

\begin{equation}
   n_{BG}^{ARCA}= \int_{\Delta \Omega} \int_{E_1}^{E_2} 2 T \frac{d \Phi_{\nu}}{dE_{\nu} d \Omega} A(E_{\nu})V(\delta)  dE_{\nu} d \Omega 
\end{equation}

where $\Delta \Omega$ is the aperture of the search cone around the source position, due to the detector Point Spread Function, dictated by the detector angular resolution. 
In this computation we consider an opening angle $\alpha=\pm 5^o$ 
around the source. This is a conservative choice that could be improved by the experiments with the better knowledge of the detector.


\begin{longtable}[c]{| c | c | c | c | c | c | c | c |}

\caption{Number of events expected to be collected in T=1 year of data taking by IceCube and KM3NeT/ARCA neutrino telescopes. We report the significance $\sigma$ that can be obtained with the number of signal $n_{events}$ and background $n_{bg}$ events collected in 1 year of data taking}.

\label{table:nevents} \\
\hline
\multicolumn{8}{|c|}{Begin of Table \ref{table:nevents}}\\
\hline
\textit{Source} & $\delta$ & $n_{events}^{IC}$ & $n_{BG}^{IC}$ & $\sigma^{IC}$ & $n_{events}^{ARCA}$  & $n_{BG}^{ARCA}$  & $\sigma^{ARCA}$  \\
\textit{Name}&  [degree] & [$1-10^3$ TeV] & [$1-10^3$ TeV] & [$1-10^3$ TeV]  & [$1-10^3$ TeV] & [$1-10^3$ TeV] & [$1-10^3$ TeV]\\

\hline
\endfirsthead

\hline
\multicolumn{8}{|c|}{Continuation of Table \ref{table:nevents}}\\
\hline
\textit{Source} & $\delta$ & $n_{events}^{IC}$ & $n_{BG}^{IC}$ & $\sigma^{IC}$ & $n_{events}^{ARCA}$  & $n_{BG}^{ARCA}$  & $\sigma^{ARCA}$  \\
\textit{Name}&  [degree] & [$1-10^3$ TeV] & [$1-10^3$ TeV] & [$1-10^3$ TeV]  & [$1-10^3$ TeV] & [$1-10^3$ TeV] & [$1-10^3$ TeV]\\
\hline
\endhead

\hline
\endfoot

\hline

\hline
\endlastfoot

\hline
VelaX-1 & -45.66 &  1.7 & $1.5 \cdot 10^{-2}$ & 14.3 & 45.4 & 40.7 & 7.1\\
HESSJ1026-582 & -58.23 &  1.0 & $1.5 \cdot 10^{-2}$ & 8.4 & 26.1 & 46.9 & 3.8\\
Crab & 22.01 &  9.07 & 34.0 & 1.6 & 6.2 & 22.7 & 1.3\\
ARGOJ2031+4157 & 42.5 &  8.0 & 35.6 & 1.34 & 4.5 & 18 & 1.1\\
HESSJ1825-137 & -13.97 &  2.7 & 4.8 & 1.3 & 11.9 & 29.9 & 2.2\\
HESSJ1718-385 & -38.51 &  0.1 & $1.5 \cdot 10^{-2}$ & 0.9 & 2.2 & 39.6 & 0.4\\
HESSJ1616-508 & -50.86 &  0.1 & $1.5 \cdot 10^{-2}$ & 0.9 & 9.5 & 46.9 & 1.4\\
MSH15-52 & -59.16 &  0.1 & $1.5 \cdot 10^{-2}$ & 0.8 & 7.2 & 46.9 & 1.1\\
TeVJ2032+4130 & 41.58 &  4.7 & 35.6 & 0.8 & 2.5 & 18.4 & 0.6\\
HESSJ1356-645 & -64.51 &  0.1 & $1.8 \cdot 10^{-2}$ & 0.8 & 6.6 & 46.9 & 0.9\\
HESSJ1708-443 & -44.29 &  0.1 & $1.5 \cdot 10^{-2}$ & 0.7 & 5.1 & 44.9 & 0.8\\
HESSJ1809-193 & -19.33 &  1.4 & 4.8 & 0.7 & 6.3 & 31.1 & 1.1\\
HESSJ1837-069 & -6.96 &  1.2 & 4.8 & 0.6 & 5.7 & 28.6 & 1.1\\
Kookaburra(PWN) & -60.75 & $6.5 \cdot 10^{-2}$ & $1.8 \cdot 10^{-2}$ & 0.5 & 4.2 & 46.9 & 0.6\\
HESSJ1303-631 & -63.19 &  0.1 & $1.8 \cdot 10^{-2}$ & 0.5 & 5.7 & 46.9 & 0.8\\
SS433e1 & 4.93 &  2.2 & 34.0 & 0.4 & 1.7 & 26.3 & 0.3\\
Boomerang & 61.24 &  2.0 & 31.6 & 0.3 & 0.5 & 6.6 & 0.2\\
HESSJ1634-472 & -47.23 &  $3.9 \cdot 10^{-2}$ & $1.5 \cdot 10^{-2}$ & 0.3 & 3.5 & 46.9 & 0.5\\
HESSJ1813-178 & -17.83 &  0.7 & 4.8 & 0.3 & 2.5 & 30.7 & 0.4\\
Geminga & 17.76 &  1.3 & 34.0 & 0.2 & 0.9 & 23.8 & 0.2\\
Westerlund1 & -46.26 & $2.6 \cdot 10^{-2}$ & $1.5 \cdot 10^{-2}$ & 0.2 & 4.7 & 46.9 & 0.7\\
Westerlund2 & -57.78 & $2.6 \cdot 10^{-2}$ & $1.5 \cdot 10^{-2}$ & 0.2 & 2.6 & 46.9 & 0.4\\
HESSJ1731-347 & -34.76 &  $2.6 \cdot 10^{-2}$ & $1.5 \cdot 10^{-2}$ & 0.2 & 1.9 & 37.0 & 0.3\\
HESSJ1826-130 & -13.02 &  0.5 & 4.8 & 0.2 & 1.5 & 29.7 & 0.3\\
HESSJ1804-216 & -21.74 &  0.4 & 4.8 & 0.2 & 2.4 & 31.7 & 0.4\\
SNRG318.2+00.1 & -59.46 &  $2.2 \cdot 10^{-2}$ & $1.5 \cdot 10^{-2}$ & 0.2 & 3.6 & 46.9 & 0.5\\
HESSJ1800-240A & -24.02 &  0.4 & 4.8 & 0.2 & 1.8 & 32.4 & 0.3\\
IGRJ18490-0000 & -0.04 &  0.3 & 4.8 & 0.1 & 0.9 & 27.3 & 0.2\\
SNRG106.3+02.7 & 60.88 &  0.7 & 31.6 & 0.1 & 0.2 & 6.8 & 0.1\\
CTA1 & 72.98 &  0.6 & 31.6 & 0.1 & 0 & 0 & - \\
HESSJ1834-087 & -8.75 &  0.2 & 4.8 & 0.1 & 1.1 & 28.9 & 0.2\\
SS433w1 & 5.04 &  0.5 & 34.0 & 0.1 & 0.4 & 26.3 & 0.1\\
HESSJ1640-465 & -46.57 &  $1.0 \cdot 10^{-2}$ & $1.5 \cdot 10^{-2}$ & 0.1 & 2 & 46.9 & 0.3\\
LS5039 & -14.85 &  0.1 & 4.8 & 0.1 & 0.6 & 30.1 & 0.1\\
W51 & 14.14 &  0.4 & 34.0 & 0.1 & 0.3 & 24.4 & 0.1\\
PSRJ2032+4127 & 41.46 &  0.4 & 35.6 & 0.1 & 0.2 & 18.4 & $4.9 \cdot 10^{-2}$\\
SNRG327.1-01.1 & -55.09 &  $7.1 \cdot 10^{-3}$ & $1.5 \cdot 10^{-2}$ & 0.1 & 0.4 & 46.9 & $6.4 \cdot 10^{-2}$\\
HESSJ1729-345 & -34.53 &  $6.4 \cdot 10^{-3}$ & $1.5 \cdot 10^{-2}$ & 0.1 & 0.6 & 36.9 & 0.1\\
CygnusX-1 & 35.20 &  0.3 & 35.6 & 0.1 & 0.18 & 19.9 & $4.0 \cdot 10^{-2}$\\
HESSJ1848-018 & -1.89 &  0.1 & 4.8 & 0.1 & 0.5 & 27.6 & 0.1\\
SNRG000.9+00.1 & -28.15 &  0.1 & 4.8 & $4.0 \cdot 10^{-2}$ & 0.4 & 33.8 & 0.1\\
LSI+61303 & 61.27 &  0.2 & 34.0 & $3.9 \cdot 10^{-2}$ & 0.2 & 27.1 & $3.5 \cdot 10^{-2}$ \\
HESSJ0632+057 & 5.79 &  0.2 & 34.0 & $3.5 \cdot 10^{-2}$ & 0.2 & 26.3 & $3.2 \cdot 10^{-2}$\\
SN1006NE & -41.8 &  $3.9 \cdot 10^{-3}$ & $1.4 \cdot 10^{-2}$ & $3.2 \cdot 10^{-2}$ & 0.3 & 41.6 & 0.1\\
HESSJ1846-029 & -2.97 &  $6.8 \cdot 10^{-2}$ & 4.8 & $3.1 \cdot 10^{-2}$  & 0.3 & 27.8 & 0.1\\
SNRG054.1+00.3 & 18.84 &  0.2 & 34.0 & $2.9 \cdot 10^{-2}$ & 0.1 & 23.4 & $2.5 \cdot 10^{-2}$ \\
TCrB & 25.92 &  0.2 & 34.0 & $2.9 \cdot 10^{-2}$ & 0.1 & 22.1 & $2.3 \cdot 10^{-2}$ \\
SNRG292.2-00.5 & -61.45 &  $2.8 \cdot 10^{-3}$ & $1.8 \cdot 10^{-2}$ & $2.1 \cdot 10^{-2}$& 0.6 & 46.9 & 0.1\\
SNRG015.4+00.1 & -15.47 & $4.1 \cdot 10^{-2}$ & 4.8 & $1.8 \cdot 10^{-2}$  & 0.2 & 30.2 & $3.1 \cdot 10^{-2}$ \\
W28 & -23.34 & $3.9 \cdot 10^{-2}$ & 4.8 & $1.8 \cdot 10^{-2}$ & 0.2 & 32.2 & $3.8 \cdot 10^{-2}$ \\
HESSJ1833-105 & -10.57 & $3.6 \cdot 10^{-2}$ & 4.8 & $1.6 \cdot 10^{-2}$ & 0.2 & 29.2 & $2.9 \cdot 10^{-2}$\\
IC443 & 22.53 & $8.7 \cdot 10^{-2}$ & 34 & $1.5 \cdot 10^{-2}$ & $5.9 \cdot 10^{-2}$ & 22.6 & $1.2 \cdot 10^{-2}$\\
3C58 & 64.85 &  $7.7 \cdot 10^{-2}$ & 31.6 & $1.4 \cdot 10^{-2}$ & $6.5 \cdot 10^{-3}$ & 2.1 & $4.5 \cdot 10^{-3}$ \\
PSRB1259-63 & -63.85 & $1.5 \cdot 10^{-3}$ & $1.8 \cdot 10^{-2}$ & $1.1 \cdot 10^{-2}$ & 0.3 & 46.9 & $4.4 \cdot 10^{-2}$ \\
CTB37A & -38.52 &  $1.0 \cdot 10^{-3}$ & $1.5 \cdot 10^{-2}$ & $8.6 \cdot 10^{-3}$ & 0.1& 39.6 & $2.3 \cdot 10^{-2}$ \\
CTB37B & -38.22 &  $9.5 \cdot 10^{-4}$ & $1.5 \cdot 10^{-2}$ & $7.9 \cdot 10^{-3}$ & 0.3 & 39.3 & $4.4 \cdot 10^{-2}$ \\
W49B & 9.09 & $4.5 \cdot 10^{-2}$ & 34.0 & $7.7 \cdot 10^{-3}$ & $3.5 \cdot 10^{-2}$ & 26.4 & $6.8 \cdot 10^{-3}$ \\
Tycho & 64.14 &  $3.7 \cdot 10^{-2}$ & 31.6 & $6.5 \cdot 10^{-3}$ & $2.3 \cdot 10^{-3}$ & 1.84 & $1.7 \cdot 10^{-3}$ \\
SNRG349.7+00.2 & -37.44 & $1.7 \cdot 10^{-4}$ & $1.5 \cdot 10^{-2}$ & $1.4 \cdot 10^{-3}$ & $6.1 \cdot 10^{-2}$ & 38.8 & $9.7 \cdot 10^{-3}$ \\
RSOphiuchi & -7 & $1.2 \cdot 10^{-3}$ & 4.8 & $5.4 \cdot 10^{-4}$ & $7.0 \cdot 10^{-3}$ & 28.5 & $1.3 \cdot 10^{-3}$ \\
\hline

\end{longtable}

For each source \textit{i} we evaluate, for IceCube and KM3NeT/ARCA, the statistical significance for neutrino detection as:

\begin{equation} \label{sigma}
    \sigma_i= \frac{n_{events,i}}{\sqrt{n_{BG,i}}}
\end{equation}

As we can see from Table \ref{table:nevents}, there are 2 sources
(Vela X and HESS J1026-582) for which the expected number of neutrino signal events in the KM3NeT/ARCA telescope is $>3$ and the $\sigma_i>3$. Both the sources in principle could have been already detected by Icecube since the $\sigma_i>7$ but the expected number of events is very low. 
In the future, KM3NeT will be able to detect or to put strong constraints on the hadronic component of these brightest sources.

As can be seen from Table \ref{table:nevents}, neutrino detection from a single source is very challenging. The neutrino detection capability can be greatly increased by virtually overlapping several sources in the same position (stacking) \citep{art:zegarelli}. Selecting events in an angular cone around the virtual source, obtained by the stacking, allows the signal to stand out with respect to the fluctuation of the background. Thanks to the stacking procedure, the cumulative significance results higher since Poissonian fluctuations of the total background are smaller than the sum of background events expected from each source. 

We evaluate the total astrophysical neutrino events as:

\begin{equation} 
n_{events}^{tot}= \sum_{i=1}^{M} n_{events}^i 
\end{equation} 

where M is the number of neutrino sources taken into account. The cumulative background is given by the sum of the single sources background in the following way:

\begin{equation} 
n_{BG}^{tot}= \sum_{i=1}^{M} n_{BG}^i 
\end{equation}

\begin{figure}[h]
    \centering
    \includegraphics[scale=0.5]{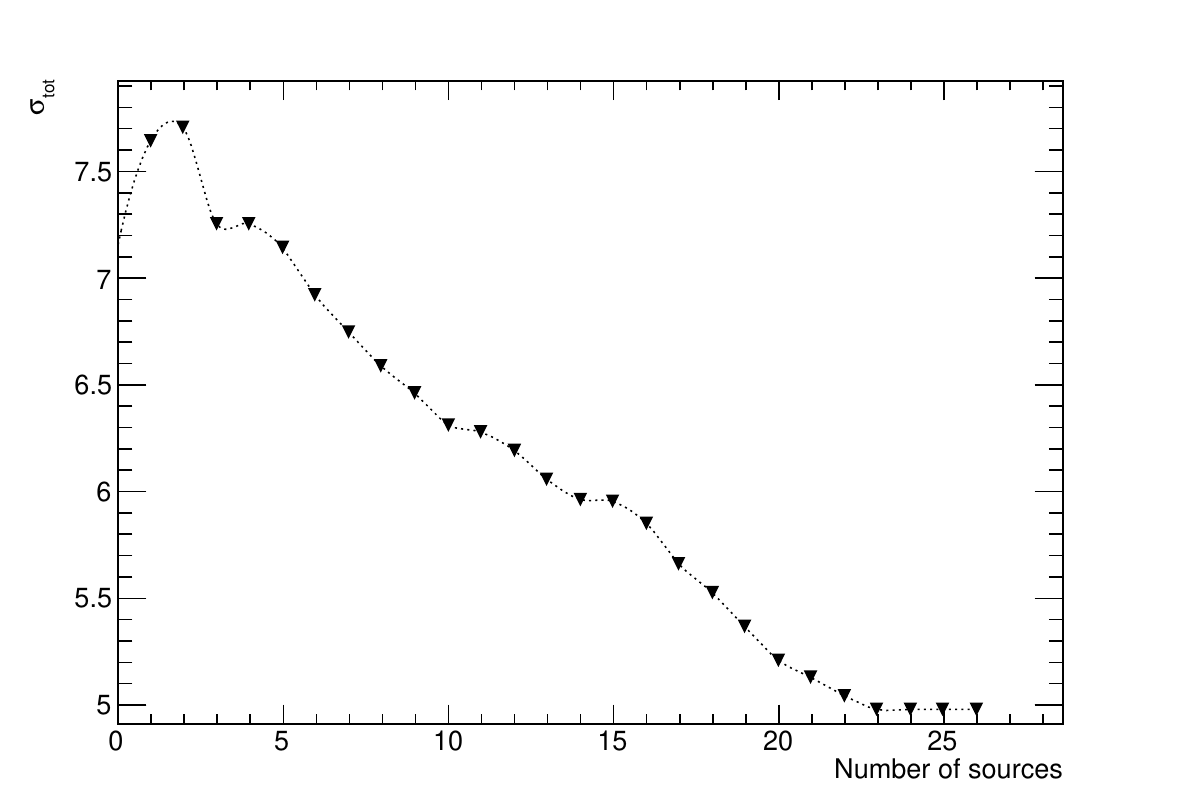}
    \caption{Stacking of PWN and PWN TeV/Halo. On the y axis the total significance computed as in Eq. \ref{eq:significativity_tot}, on the x axis the number of stacked sources.}
    \label{fig:pwn_arca}
\end{figure}

We perform a stacking analysis on a single category of sources, in particular we selected a sample containing M=25 PWN and PWN/Halo since they are the ones with the highest level of significance in Table \ref{table:nevents}. We list the 25 sources in decreasing order of significance. We evaluate how the statistical significance ($\sigma_{tot}(m)$) of the result obtained by the stacking changes with the number m of the sources included in the evaluation, 

\begin{equation} 
\sigma_{tot}(m)= \frac{\sum_{i=1}^m n_{events,i}}{\sqrt{\sum_{i=1}^m n_{BG,i}}} 
\label{eq:significativity_tot}
\end{equation} 

where m = (2, 3, 4, 5, ...M) is the number of overlapped galactic sources. In Figure \ref{fig:pwn_arca} we plot the variation of $\sigma_{tot}(m)$: we find that with data sample accumulated by KM3NeT/ARCA telescope in 1 year, by stacking $\sim 3$ sources, we achieved a significance value $\sigma \sim 7.5$. We apply the same stacking technique to other categories of sources, as SNR/Composite and Shell. As a result we find that, even summing up all the sources, the signal is far below the detector sensitivity. The stacking technique can be applied also to ANTARES and IceCube detectors that collected data for several years. 

We estimate the telescope horizon (the maximal distance at which a source can be detected) considering different classes of sources of our sample, i.e. PWNe, SNRs, MQs, Novae, and we select, as reference, the source with the highest significance for each type. For these sources we know $dN_{Ref}/dE$, $d_{Ref}$, and $n_{BG, Ref}$. We indirectly define the horizon distance as:

\begin{equation} \sigma= \frac{n_{events}}{\sqrt{n_{BG,Ref}}} = \frac{T}{\sqrt{n_{BG,Ref} }} \frac{d_{Ref}^2}{d_{horizon}^2} \int \frac{dN_{Ref}}{dE dS} A(E_{\nu}) dE_{\nu}  
\end{equation}

\begin{figure}[ht]
    \centering
    \includegraphics[scale=0.5]{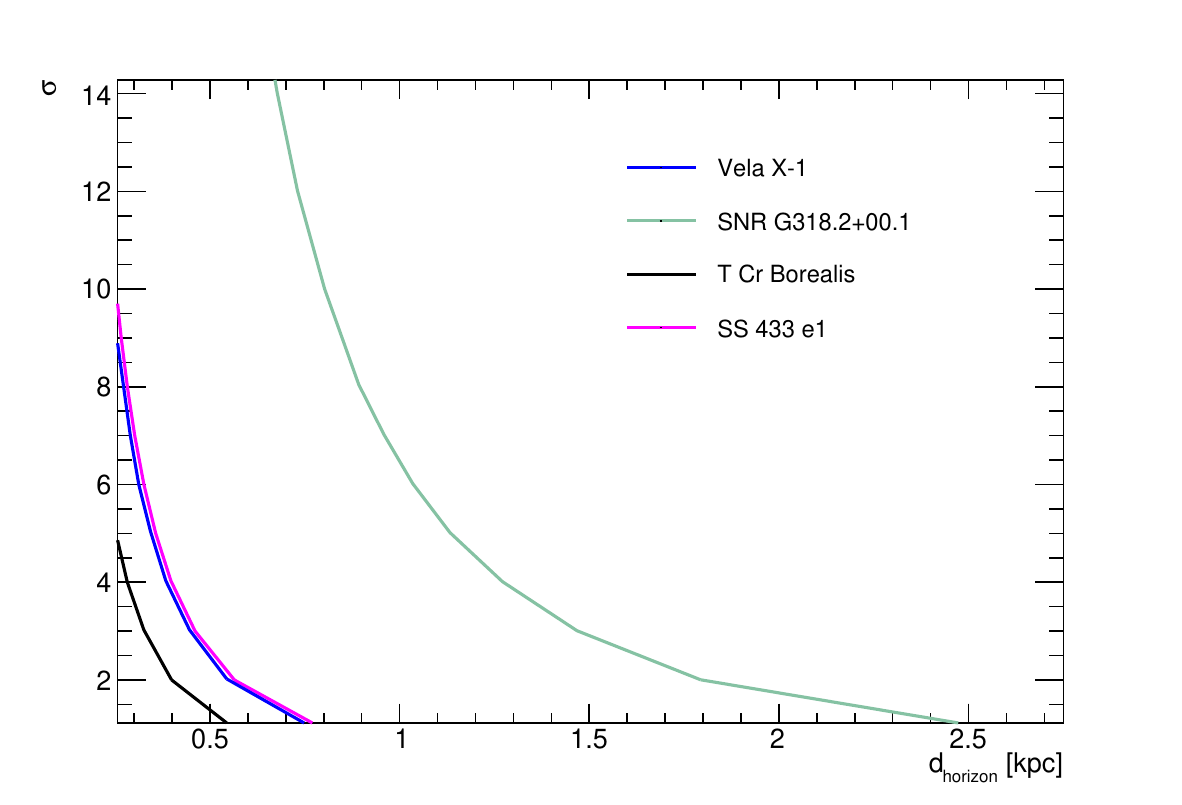}
    \caption{KM3NeT sensitivity as function of the distance in kpc evaluated for different classes of sources: PWNs, SNRs, MQs, Novae. For each class we selected the source with the higher sensitivity $\sigma$ (see Table \ref{table:nevents}), i.e. Vela X-1 for the PWN class, SNR G318.2+00.1 for the SNR class, SS 433 e1 for the MQ class and TCrBorealis for the Novae.}
    \label{fig:sigma_d}
\end{figure}

Figure \ref{fig:sigma_d} shows the possibility to detect in 1 year, for a detector like KM3NeT/ARCA, the different types of point like sources as a function of the distance and of the statistical significance required.



\section{Discussion and Conclusions}\label{sec:discussion}

In this paper we estimate the neutrino flux originated in galactic point like sources from the observed high energy photon flux. We assume that the entire TeV $\gamma$-ray flux is due to the hadronic component. This assumption permit us to estimate the maximum neutrino flux expected from each galactic source. A non detection of the expected neutrino flux will allow to constraint the hadronic component. We select all the galactic sources that have been detected at TeV energies and we compare the total neutrino flux with the diffuse galactic flux reported by IceCube.
We also predict the number of neutrino events expected to be detected from individual galactic sources by present and future neutrinos telescopes.  
We obtain a number of interesting results, specified below: 

\begin{enumerate}

\item
If the entire TeV emission is due to the hadronic component, the galactic sources can contribute to the diffuse neutrino flux detected by IceCube at the level of$\sim 20\%$ up to $\sim 10 $ TeV and at $>50\%$ at higher energies. Considering the fact that part of the TeV emission could also be caused by leptonic mechanism inside the sources, our results may be considered as an upper limit on the hadronic contribution to the high energy photon emission. 
\item
Even if all the TeV emission is due to the hadronic processes, no galactic source is detectable by IceCube in 1 year of data taking. Some sources may be detectable by KM3NeT/ARCA in 1 year, in particular, the most promising ones result to be PWN type, i.e. Vela X-1 with a significance $\sigma \sim 7$ and HESS J1026-582  with a significance $\sigma \sim 4$. 
Therefore, if KM3NeT/ARCA will not detect neutrino from these promising sources, it will be possible to constraint the hadronic component responsible of the $\gamma$-ray high energy emission.
\item 
It is important to note that the estimate of the number of expected neutrinos from the galactic sources has been performed without considering any specific physical model for the source to produce the TeV fluxes. Our analysis can be performed also to put strong constraints on some specific models for high energy neutrino emission. 
\item Applying the stacking procedure to the most promising galactic neutrino sources, PWN/PWN TeV Halo, according to our calculation, in 1 year KM3NeT should be able to detect (with $\sigma \sim 7.5$ statistical significance) a neutrino flux from the stacking of at least 3 PWN/PWN TeV Halo. 
\item We estimate the maximum distance at which the most luminous source (SNR G318.2+00.1) in our catalogue
can be detected in 1 year by KM3NeT/ARCA and we find that $d_{horizon}< 3$ kpc. The detection significance vs distance relation is only based on the source luminosity. 

\end{enumerate}

To conclude, we find that several sources may be strong emitters of high energy neutrinos. Future investigations are needed to determine which are the sources responsible of IceCube diffuse flux and to constrain the hadronic component responsible to the $\gamma$-ray TeV emission.\\


\section{Acknowledgement}
We thank Angela Zegarelli and Massimo Della Valle for their suggestions.
We also thank Bradley Schaefer
for having suggested the neutrino estimate from the NOVA T CrB.
\clearpage

\bibliography{biblio.bib}{}

\begin{thebibliography}{}
\expandafter\ifx\csname natexlab\endcsname\relax\def\natexlab#1{#1}\fi
\providecommand{\url}[1]{\href{#1}{#1}}
\providecommand{\dodoi}[1]{doi:~\href{http://doi.org/#1}{\nolinkurl{#1}}}
\providecommand{\doeprint}[1]{\href{http://ascl.net/#1}{\nolinkurl{http://ascl.net/#1}}}
\providecommand{\doarXiv}[1]{\href{https://arxiv.org/abs/#1}{\nolinkurl{https://arxiv.org/abs/#1}}}

\bibitem[{Aartsen {et~al.}(2017)Aartsen, Ackermann, Adams, Aguilar, Ahlers, Ahrens, Samarai, Altmann, Andeen, Anderson, Ansseau, Anton, Argüelles, Auffenberg, Axani, Bagherpour, Bai, Barron, Barwick, Baum, Bay, Beatty, Tjus, Becker, BenZvi, Berley, Bernardini, Besson, Binder, Bindig, Blaufuss, Blot, Bohm, Börner, Bos, Bose, Böser, Botner, Bourbeau, Bradascio, Braun, Brayeur, Brenzke, Bretz, Bron, Burgman, Carver, Casey, Casier, Cheung, Chirkin, Christov, Clark, Classen, Coenders, Collin, Conrad, Cowen, Cross, Day, de~André, Clercq, DeLaunay, Dembinski, Ridder, Desiati, de~Vries, de~Wasseige, de~With, DeYoung, Díaz-Vélez, di~Lorenzo, Dujmovic, Dumm, Dunkman, Eberhardt, Ehrhardt, Eichmann, Eller, Evenson, Fahey, Fazely, Felde, Filimonov, Finley, Flis, Franckowiak, Friedman, Fuchs, Gaisser, Gallagher, Gerhardt, Ghorbani, Giang, Glauch, Glüsenkamp, Goldschmidt, Gonzalez, Grant, Griffith, Haack, Hallgren, Halzen, Hanson, Hebecker, Heereman, Helbing, Hellauer, Hickford, Hignight, Hill, Hoffman, Hoffmann,
  Hokanson-Fasig, Hoshina, Huang, Huber, Hultqvist, In, Ishihara, Jacobi, Japaridze, Jeong, Jero, Jones, Kalacynski, Kang, Kappes, Karg, Karle, Katz, Kauer, Keivani, Kelley, Kheirandish, Kim, Kim, Kintscher, Kiryluk, Kittler, Klein, Kohnen, Koirala, Kolanoski, Köpke, Kopper, Kopper, Koschinsky, Koskinen, Kowalski, Krings, Kroll, Krückl, Kunnen, Kunwar, Kurahashi, Kuwabara, Kyriacou, Labare, Lanfranchi, Larson, Lauber, Lennarz, Lesiak-Bzdak, Leuermann, Liu, Lu, Lünemann, Luszczak, Madsen, Maggi, Mahn, Mancina, Maruyama, Mase, Maunu, McNally, Meagher, Medici, Meier, Menne, Merino, Meures, Miarecki, Micallef, Momenté, Montaruli, Moore, Moulai, Nahnhauer, Nakarmi, Naumann, Neer, Niederhausen, Nowicki, Nygren, Pollmann, Olivas, O’Murchadha, Palczewski, Pandya, Pankova, Peiffer, Pepper, de~los Heros, Pieloth, Pinat, Plum, Price, Przybylski, Raab, Rädel, Rameez, Rawlins, Reimann, Relethford, Relich, Resconi, Rhode, Richman, Robertson, Rongen, Rott, Ruhe, Ryckbosch, Rysewyk, Sälzer, Herrera, Sandrock,
  Sandroos, Sarkar, Sarkar, Satalecka, Schlunder, Schmidt, Schneider, Schoenen, Schöneberg, Schumacher, Seckel, Seunarine, Soldin, Song, Spiczak, Spiering, Stachurska, Stanev, Stasik, Stettner, Steuer, Stezelberger, Stokstad, Stößl, Strotjohann, Sullivan, Sutherland, Taboada, Tatar, Tenholt, Ter-Antonyan, Terliuk, Tešić, Tilav, Toale, Tobin, Toscano, Tosi, Tselengidou, Tung, Turcati, Turley, Ty, Unger, Usner, Vandenbroucke, Driessche, van Eijndhoven, Vanheule, van Santen, Vehring, Vogel, Vraeghe, Walck, Wallace, Wallraff, Wandler, Wandkowsky, Waza, Weaver, Weiss, Wendt, Westerhoff, Whelan, Wickmann, Wiebe, Wiebusch, Wille, Williams, Wills, Wolf, Wood, Wood, Woolsey, Woschnagg, Xu, Xu, Xu, Yanez, Yodh, Yoshida, Yuan, Zoll, \& Collaboration}]{artsen2017}
Aartsen, M.~G., Ackermann, M., Adams, J., {et~al.} 2017, The Astrophysical Journal, 849, 67, \dodoi{10.3847/1538-4357/aa8dfb}

\bibitem[{Aartsen(2014)}]{art:icecube_aeff}
Aartsen, M. G. e. a.~C. 2014, The Astrophysical Journal, 796, \dodoi{10.1088/0004-637X/796/2/109}

\bibitem[{{Abbasi} {et~al.}(2011){Abbasi}, {Abdou}, {Abu-Zayyad}, {Adams}, {Aguilar}, {Ahlers}, {Altmann}, {Andeen}, {Auffenberg}, {Bai}, {Baker}, {Barwick}, {Bay}, {Bazo Alba}, {Beattie}, {Beatty}, {Bechet}, {Becker}, {Becker}, {Benabderrahmane}, {Benzvi}, {Berdermann}, {Berghaus}, {Berley}, {Bernardini}, {Bertrand}, {Besson}, {Bindig}, {Bissok}, {Blaufuss}, {Blumenthal}, {Boersma}, {Bohm}, {Bose}, {B{\"o}ser}, {Botner}, {Brown}, {Buitink}, {Caballero-Mora}, {Carson}, {Chirkin}, {Christy}, {Clem}, {Clevermann}, {Cohen}, {Colnard}, {Cowen}, {D'Agostino}, {Danninger}, {Daughhetee}, {Davis}, {de Clercq}, {Demir{\"o}rs}, {Denger}, {Depaepe}, {Descamps}, {Desiati}, {de Vries-Uiterweerd}, {Deyoung}, {D{\'\i}az-V{\'e}lez}, {Dierckxsens}, {Dreyer}, {Dumm}, {Ehrlich}, {Eisch}, {Ellsworth}, {Engdeg{\r{a}}rd}, {Euler}, {Evenson}, {Fadiran}, {Fazely}, {Fedynitch}, {Feintzeig}, {Feusels}, {Filimonov}, {Finley}, {Fischer-Wasels}, {Foerster}, {Fox}, {Franckowiak}, {Franke}, {Gaisser}, {Gallagher}, {Gerhardt}, {Gladstone},
  {Gl{\"u}senkamp}, {Goldschmidt}, {Goodman}, {Gora}, {Grant}, {Griesel}, {Gro{\ss}}, {Grullon}, {Gurtner}, {Ha}, {Hajismail}, {Hallgren}, {Halzen}, {Han}, {Hanson}, {Heinen}, {Helbing}, {Herquet}, {Hickford}, {Hill}, {Hoffman}, {Homeier}, {Hoshina}, {Hubert}, {Huelsnitz}, {H{\"u}l{\ss}}, {Hulth}, {Hultqvist}, {Hussain}, {Ishihara}, {Jacobsen}, {Japaridze}, {Johansson}, {Joseph}, {Kampert}, {Kappes}, {Karg}, {Karle}, {Kenny}, {Kiryluk}, {Kislat}, {Klein}, {K{\"o}hne}, {Kohnen}, {Kolanoski}, {K{\"o}pke}, {Kopper}, {Koskinen}, {Kowalski}, {Kowarik}, {Krasberg}, {Krings}, {Kroll}, {Kurahashi}, {Kuwabara}, {Labare}, {Lafebre}, {Laihem}, {Landsman}, {Larson}, {Lauer}, {L{\"u}nemann}, {Madsen}, {Majumdar}, {Marotta}, {Maruyama}, {Mase}, {Matis}, {Meagher}, {Merck}, {M{\'e}sz{\'a}ros}, {Meures}, {Middell}, {Milke}, {Miller}, {Montaruli}, {Morse}, {Movit}, {Nahnhauer}, {Nam}, {Naumann}, {Nie{\ss}en}, {Nygren}, {Odrowski}, {Olivas}, {Olivo}, {O'Murchadha}, {Ono}, {Panknin}, {Paul}, {P{\'e}rez de Los Heros},
  {Petrovic}, {Piegsa}, {Pieloth}, {Porrata}, {Posselt}, {Price}, {Przybylski}, {Rawlins}, {Redl}, {Resconi}, {Rhode}, {Ribordy}, {Rizzo}, {Rodrigues}, {Roth}, {Rothmaier}, {Rott}, {Ruhe}, {Rutledge}, {Ruzybayev}, {Ryckbosch}, {Sander}, {Santander}, {Sarkar}, {Schatto}, {Schmidt}, {Sch{\"o}nwald}, {Schukraft}, {Schultes}, {Schulz}, {Schunck}, {Seckel}, {Semburg}, {Seo}, {Sestayo}, {Seunarine}, {Silvestri}, {Slipak}, {Spiczak}, {Spiering}, {Stamatikos}, {Stanev}, {Stephens}, {Stezelberger}, {Stokstad}, {St{\"o}ssl}, {Stoyanov}, {Strahler}, {Straszheim}, {St{\"u}r}, {Sullivan}, {Swillens}, {Taavola}, {Taboada}, {Tamburro}, {Tepe}, {Ter-Antonyan}, {Tilav}, {Toale}, {Toscano}, {Tosi}, {Tur{\v{c}}an}, {van Eijndhoven}, {Vandenbroucke}, {van Overloop}, {van Santen}, {Vehring}, {Voge}, {Walck}, {Waldenmaier}, {Wallraff}, {Walter}, {Weaver}, {Wendt}, {Westerhoff}, {Whitehorn}, {Wiebe}, {Wiebusch}, {Williams}, {Wischnewski}, {Wissing}, {Wolf}, {Wood}, {Woschnagg}, {Xu}, {Xu}, {Yodh}, {Yoshida}, {Zarzhitsky}, \&
  {Zoll}}]{art:atmosferici}
{Abbasi}, R., {Abdou}, Y., {Abu-Zayyad}, T., {et~al.} 2011, \prd, 84, 082001, \dodoi{10.1103/PhysRevD.84.082001}

\bibitem[{Abbasi {et~al.}(2023)Abbasi, Ackermann, Adams, Aguilar, Ahlers, Ahrens, Alameddine, Alves, Amin, Andeen, Anderson, Anton, Argüelles, Ashida, Athanasiadou, Axani, Bai, V., Barwick, Basu, Baur, Bay, Beatty, Becker, Tjus, Beise, Bellenghi, Benda, BenZvi, Berley, Bernardini, Besson, Binder, Bindig, Blaufuss, Blot, Boddenberg, Bontempo, Book, Borowka, Böser, Botner, Böttcher, Bourbeau, Bradascio, Braun, Brinson, Bron, Brostean-Kaiser, Burley, Busse, Campana, Carnie-Bronca, Chen, Chen, Chirkin, Choi, Clark, Clark, Classen, Coleman, Collin, Connolly, Conrad, Coppin, Correa, Cowen, Cross, Dappen, Dave, Clercq, DeLaunay, López, Dembinski, Deoskar, Desai, Desiati, de~Vries, de~Wasseige, DeYoung, Diaz, Díaz-Vélez, Dittmer, Dujmovic, Dunkman, DuVernois, Ehrhardt, Eller, Engel, Erpenbeck, Evans, Evenson, Fan, Fazely, Fedynitch, Feigl, Fiedlschuster, Fienberg, Finley, Fischer, Fox, Franckowiak, Friedman, Fritz, Fürst, Gaisser, Gallagher, Ganster, Garcia, Garrappa, Gerhardt, Ghadimi, Glaser, Glauch,
  Glüsenkamp, Goehlke, Goldschmidt, Gonzalez, Goswami, Grant, Grégoire, Griswold, Günther, Gutjahr, Haack, Hallgren, Halliday, Halve, Halzen, Minh, Hanson, Hardin, Harnisch, Haungs, Helbing, Henningsen, Hettinger, Hickford, Hignight, Hill, Hill, Hoffman, Hoshina, Hou, Huang, Huber, Huber, Hultqvist, Hünnefeld, Hussain, Hymon, In, Iovine, Ishihara, Jansson, Japaridze, Jeong, Jin, Jones, Kang, Kang, Kang, Kappes, Kappesser, Kardum, Karg, Karl, Karle, Katz, Kauer, Kellermann, Kelley, Kheirandish, Kin, Kiryluk, Klein, Kochocki, Koirala, Kolanoski, Kontrimas, Köpke, Kopper, Kopper, Koskinen, Koundal, Kovacevich, Kowalski, Kozynets, Krupczak, Kun, Kurahashi, Lad, Gualda, Lanfranchi, Larson, Lauber, Lazar, Lee, Leonard, Leszczyńska, Li, Lincetto, Liu, Liubarska, Lohfink, Mariscal, Lu, Lucarelli, Ludwig, Luszczak, Lyu, Ma, Madsen, Mahn, Makino, Mancina, Mariş, Martinez-Soler, Maruyama, McCarthy, McElroy, McNally, Mead, Meagher, Mechbal, Medina, Meier, Meighen-Berger, Merckx, Micallef, Mockler, Montaruli,
  Moore, Morik, Morse, Moulai, Mukherjee, Naab, Nagai, Nahnhauer, Naumann, Necker, Nguyen, Niederhausen, Nisa, Nowicki, Nygren, Pollmann, Oehler, Oeyen, Olivas, O'Sullivan, Pandya, Pankova, Park, Parker, Paudel, Paul, de~los Heros, Peters, Peterson, Philippen, Pieper, Pizzuto, Plum, Popovych, Porcelli, Rodriguez, Pries, Przybylski, Raab, Rack-Helleis, Raissi, Rameez, Rawlins, Rea, Rechav, Rehman, Reichherzer, Reimann, Renzi, Resconi, Reusch, Rhode, Richman, Riedel, Roberts, Robertson, Roellinghoff, Rongen, Rott, Ruhe, Ryckbosch, Cantu, Safa, Saffer, Salazar-Gallegos, Sampathkumar, Herrera, Sandrock, Santander, Sarkar, Sarkar, Satalecka, Schaufel, Schieler, Schindler, Schmidt, Schneider, Schneider, Schröder, Schumacher, Schwefer, Sclafani, Seckel, Seunarine, Sharma, Shefali, Shimizu, Silva, Skrzypek, Smithers, Snihur, Soedingrekso, Sogaard, Soldin, Spannfellner, Spiczak, Spiering, Stamatikos, Stanev, Stein, Stettner, Stezelberger, Stokstad, Stürwald, Stuttard, Sullivan, Taboada, Ter-Antonyan, Thwaites,
  Tilav, Tischbein, Tollefson, Tönnis, Toscano, Tosi, Trettin, Tselengidou, Tung, Turcati, Turcotte, Turley, Twagirayezu, Ty, Elorrieta, Valtonen-Mattila, Vandenbroucke, van Eijndhoven, Vannerom, van Santen, Veitch-Michaelis, Verpoest, Walck, Wang, Watson, Weaver, Weigel, Weindl, Weiss, Weldert, Wendt, Werthebach, Weyrauch, Whitehorn, Wiebusch, Willey, Williams, Wolf, Wrede, Wulff, Xu, Yanez, Yildizci, Yoshida, Yu, Yuan, Zhang, \& Zhelnin}]{icecube_2023}
Abbasi, R., Ackermann, M., Adams, J., {et~al.} 2023, Science, 380, 1338, \dodoi{10.1126/science.adc9818}

\bibitem[{{Abdo} {et~al.}(2009){Abdo}, {Allen}, {Aune}, {Berley}, {Chen}, {Christopher}, {DeYoung}, {Dingus}, {Ellsworth}, {Gonzalez}, {Goodman}, {Hays}, {Hoffman}, {H{\"u}ntemeyer}, {Kolterman}, {Linnemann}, {McEnery}, {Morgan}, {Mincer}, {Nemethy}, {Pretz}, {Ryan}, {Saz Parkinson}, {Shoup}, {Sinnis}, {Smith}, {Vasileiou}, {Walker}, {Williams}, \& {Yodh}}]{Geminga}
{Abdo}, A.~A., {Allen}, B.~T., {Aune}, T., {et~al.} 2009, \apjl, 700, L127, \dodoi{10.1088/0004-637X/700/2/L127}

\bibitem[{{Abdo} {et~al.}(2012){Abdo}, {Abeysekara}, {Allen}, {Aune}, {Berley}, {Bonamente}, {Christopher}, {DeYoung}, {Dingus}, {Ellsworth}, {Galbraith-Frew}, {Gonzalez}, {Goodman}, {Hoffman}, {H{\"u}ntemeyer}, {Hui}, {Kolterman}, {Linnemann}, {McEnery}, {Mincer}, {Morgan}, {Nemethy}, {Pretz}, {Ryan}, {Saz Parkinson}, {Shoup}, {Sinnis}, {Smith}, {Vasileiou}, {Walker}, {Williams}, \& {Yodh}}]{TeVJ2032+4130}
{Abdo}, A.~A., {Abeysekara}, U., {Allen}, B.~T., {et~al.} 2012, \apj, 753, 159, \dodoi{10.1088/0004-637X/753/2/159}

\bibitem[{Abeysekara {et~al.}(2017)Abeysekara, Albert, Alfaro, Alvarez, Álvarez, Arceo, Arteaga-Velázquez, Rojas, Solares, Barber, Bautista-Elivar, Becerril, Belmont-Moreno, BenZvi, Berley, Bernal, Braun, Brisbois, Caballero-Mora, Capistrán, Carramiñana, Casanova, Castillo, Cotti, Cotzomi, de~León, León, la~Fuente, Dingus, DuVernois, Díaz-Vélez, Ellsworth, Engel, Enríquez-Rivera, Fiorino, Fraija, García-González, Garfias, Gerhardt, Muñoz, González, Goodman, Hampel-Arias, Harding, Hernández, Hernández-Almada, Hinton, Hona, Hui, Hüntemeyer, Iriarte, Jardin-Blicq, Joshi, Kaufmann, Kieda, Lara, Lauer, Lee, Lennarz, Vargas, Linnemann, Longinotti, Raya, Luna-García, López-Coto, Malone, Marinelli, Martinez, Martinez-Castellanos, Martínez-Castro, Martínez-Huerta, Matthews, Miranda-Romagnoli, Moreno, Mostafá, Nellen, Newbold, Nisa, Noriega-Papaqui, Pelayo, Pretz, Pérez-Pérez, Ren, Rho, Rivière, Rosa-González, Rosenberg, Ruiz-Velasco, Salazar, Greus, Sandoval, Schneider, Schoorlemmer, Sinnis,
  Smith, Springer, Surajbali, Taboada, Tibolla, Tollefson, Torres, Ukwatta, Vianello, Weisgarber, Westerhoff, Wisher, Wood, Yapici, Yodh, Younk, Zepeda, Zhou, Guo, Hahn, Li, \& Zhang}]{tevhalo_hawc}
Abeysekara, A.~U., Albert, A., Alfaro, R., {et~al.} 2017, Science, 358, 911, \dodoi{10.1126/science.aan4880}

\bibitem[{{Abeysekara} {et~al.}(2017){Abeysekara}, {Albert}, {Alfaro}, {Alvarez}, {{\'A}lvarez}, {Arceo}, {Arteaga-Vel{\'a}zquez}, {Ayala Solares}, {Barber}, {Baughman}, {Bautista-Elivar}, {Becerra Gonzalez}, {Becerril}, {Belmont-Moreno}, {BenZvi}, {Berley}, {Bernal}, {Braun}, {Brisbois}, {Caballero-Mora}, {Capistr{\'a}n}, {Carrami{\~n}ana}, {Casanova}, {Castillo}, {Cotti}, {Cotzomi}, {Couti{\~n}o de Le{\'o}n}, {de la Fuente}, {De Le{\'o}n}, {Diaz Hernandez}, {Dingus}, {DuVernois}, {D{\'\i}az-V{\'e}lez}, {Ellsworth}, {Engel}, {Fiorino}, {Fraija}, {Garc{\'\i}a-Gonz{\'a}lez}, {Garfias}, {Gerhardt}, {Gonz{\'a}lez Mu{\~n}oz}, {Gonz{\'a}lez}, {Goodman}, {Hampel-Arias}, {Harding}, {Hernandez}, {Hernandez-Almada}, {Hinton}, {Hui}, {H{\"u}ntemeyer}, {Iriarte}, {Jardin-Blicq}, {Joshi}, {Kaufmann}, {Kieda}, {Lara}, {Lauer}, {Lee}, {Lennarz}, {Le{\'o}n Vargas}, {Linnemann}, {Longinotti}, {Raya}, {Luna-Garc{\'\i}a}, {L{\'o}pez-Coto}, {Malone}, {Marinelli}, {Martinez}, {Martinez-Castellanos}, {Mart{\'\i}nez-Castro},
  {Mart{\'\i}nez-Huerta}, {Matthews}, {Miranda-Romagnoli}, {Moreno}, {Mostaf{\'a}}, {Nellen}, {Newbold}, {Nisa}, {Noriega-Papaqui}, {Pelayo}, {Pretz}, {P{\'e}rez-P{\'e}rez}, {Ren}, {Rho}, {Rivi{\`e}re}, {Rosa-Gonz{\'a}lez}, {Rosenberg}, {Ruiz-Velasco}, {Salazar}, {Salesa Greus}, {Sandoval}, {Schneider}, {Schoorlemmer}, {Sinnis}, {Smith}, {Springer}, {Surajbali}, {Taboada}, {Tibolla}, {Tollefson}, {Torres}, {Ukwatta}, {Vianello}, {Villase{\~n}or}, {Weisgarber}, {Westerhoff}, {Wisher}, {Wood}, {Yapici}, {Younk}, {Zepeda}, \& {Zhou}}]{art:HAWC_cat}
{Abeysekara}, A.~U., {Albert}, A., {Alfaro}, R., {et~al.} 2017, ApJ, 843, 40, \dodoi{10.3847/1538-4357/aa7556}

\bibitem[{{Abeysekara} {et~al.}(2018){Abeysekara}, {Benbow}, {Bird}, {Brill}, {Brose}, {Buckley}, {Chromey}, {Daniel}, {Falcone}, {Finley}, {Fortson}, {Furniss}, {Gent}, {Gillanders}, {Hanna}, {Hassan}, {Hervet}, {Holder}, {Hughes}, {Humensky}, {Kaaret}, {Kar}, {Kertzman}, {Kieda}, {Krause}, {Krennrich}, {Kumar}, {Lang}, {Lin}, {Maier}, {Moriarty}, {Mukherjee}, {O'Brien}, {Ong}, {Otte}, {Park}, {Petrashyk}, {Pohl}, {Pueschel}, {Quinn}, {Ragan}, {Richards}, {Roache}, {Sadeh}, {Santander}, {Schlenstedt}, {Sembroski}, {Sushch}, {Tyler}, {Vassiliev}, {Wakely}, {Weinstein}, {Wells}, {Wilcox}, {Wilhelm}, {Williams}, {Williamson}, {Zitzer}, {VERITAS Collaboration}, {Acciari}, {Ansoldi}, {Antonelli}, {Arbet Engels}, {Baack}, {Babi{\'c}}, {Banerjee}, {Barres de Almeida}, {Barrio}, {Becerra Gonz{\'a}lez}, {Bednarek}, {Bernardini}, {Berti}, {Besenrieder}, {Bhattacharyya}, {Bigongiari}, {Biland}, {Blanch}, {Bonnoli}, {Busetto}, {Carosi}, {Ceribella}, {Cikota}, {Colak}, {Colin}, {Colombo}, {Contreras}, {Cortina},
  {Covino}, {D'Elia}, {Da Vela}, {Dazzi}, {De Angelis}, {De Lotto}, {Delfino}, {Delgado}, {Di Pierro}, {Do Souto Espi{\~n}era}, {Dom{\'\i}nguez}, {Dominis Prester}, {Dorner}, {Doro}, {Einecke}, {Elsaesser}, {Fallah Ramazani}, {Fattorini}, {Fern{\'a}ndez-Barral}, {Ferrara}, {Fidalgo}, {Foffano}, {Fonseca}, {Font}, {Fruck}, {Galindo}, {Gallozzi}, {Garc{\'\i}a L{\'o}pez}, {Garczarczyk}, {Gasparyan}, {Gaug}, {Giammaria}, {Godinovi{\'c}}, {Guberman}, {Hadasch}, {Hahn}, {Herrera}, {Hoang}, {Hrupec}, {Inoue}, {Ishio}, {Iwamura}, {Kubo}, {Kushida}, {Kuve{\v{z}}di{\'c}}, {Lamastra}, {Lelas}, {Leone}, {Lindfors}, {Lombardi}, {Longo}, {L{\'o}pez}, {L{\'o}pez-Oramas}, {Machado de Oliveira Fraga}, {Maggio}, {Majumdar}, {Makariev}, {Mallamaci}, {Maneva}, {Manganaro}, {Mannheim}, {Maraschi}, {Mariotti}, {Mart{\'\i}nez}, {Masuda}, {Mazin}, {Minev}, {Miranda}, {Mirzoyan}, {Molina}, {Moralejo}, {Moreno}, {Moretti}, {Munar-Adrover}, {Neustroev}, {Niedzwiecki}, {Nievas Rosillo}, {Nigro}, {Nilsson}, {Ninci}, {Nishijima}, {Noda},
  {Nogu{\'e}s}, {N{\"o}the}, {Paiano}, {Palacio}, {Paneque}, {Paoletti}, {Paredes}, {Pedaletti}, {Pe{\~n}il}, {Peresano}, {Persic}, {Prada Moroni}, {Prandini}, {Puljak}, {Garcia}, {Rhode}, {Rib{\'o}}, {Rico}, {Righi}, {Rugliancich}, {Saha}, {Sahakyan}, {Saito}, {Satalecka}, {Schweizer}, {Sitarek}, {{\v{S}}nidari{\'c}}, {Sobczynska}, {Somero}, {Stamerra}, {Strzys}, {Suri{\'c}}, {Tavecchio}, {Temnikov}, {Terzi{\'c}}, {Teshima}, {Torres-Alb{\`a}}, {Tsujimoto}, {van Scherpenberg}, {Vanzo}, {Vazquez Acosta}, {Vovk}, {Will}, {Zari{\'c}}, \& {MAGIC Collaboration}}]{PSRJ2032+4127}
{Abeysekara}, A.~U., {Benbow}, W., {Bird}, R., {et~al.} 2018, \apjl, 867, L19, \dodoi{10.3847/2041-8213/aae70e}

\bibitem[{{Acciari} {et~al.}(2009){Acciari}, {Aliu}, {Arlen}, {Aune}, {Bautista}, {Beilicke}, {Benbow}, {Boltuch}, {Bradbury}, {Buckley}, {Bugaev}, {Butt}, {Byrum}, {Cannon}, {Cesarini}, {Chow}, {Ciupik}, {Cogan}, {Cui}, {Dickherber}, {Ergin}, {Fegan}, {Finley}, {Fortin}, {Fortson}, {Furniss}, {Gall}, {Gillanders}, {Gotthelf}, {Grube}, {Guenette}, {Gyuk}, {Hanna}, {Holder}, {Horan}, {Hui}, {Humensky}, {Kaaret}, {Karlsson}, {Kertzman}, {Kieda}, {Konopelko}, {Krawczynski}, {Krennrich}, {Lang}, {LeBohec}, {Maier}, {McCann}, {McCutcheon}, {Millis}, {Moriarty}, {Mukherjee}, {Ong}, {Otte}, {Pandel}, {Perkins}, {Pohl}, {Quinn}, {Ragan}, {Reyes}, {Reynolds}, {Roache}, {Rose}, {Schroedter}, {Sembroski}, {Smith}, {Steele}, {Swordy}, {Theiling}, {Toner}, {Vassiliev}, {Vincent}, {Wagner}, {Wakely}, {Ward}, {Weekes}, {Weinstein}, {Weisgarber}, {Williams}, {Wissel}, {Wood}, \& {Zitzer}}]{art:SNRG106.3}
{Acciari}, V.~A., {Aliu}, E., {Arlen}, T., {et~al.} 2009, ApJL, 703, L6, \dodoi{10.1088/0004-637X/703/1/L6}

\bibitem[{{Acciari} {et~al.}(2022){Acciari}, {Ansoldi}, {Antonelli}, {Arbet Engels}, {Artero}, {Asano}, {Baack}, {Babi{\'c}}, {Baquero}, {Barres de Almeida}, {Barrio}, {Batkovi{\'c}}, {Becerra Gonz{\'a}lez}, {Bednarek}, {Bellizzi}, {Bernardini}, {Bernardos}, {Berti}, {Besenrieder}, {Bhattacharyya}, {Bigongiari}, {Biland}, {Blanch}, {B{\"o}kenkamp}, {Bonnoli}, {Bo{\v{s}}njak}, {Busetto}, {Carosi}, {Ceribella}, {Cerruti}, {Chai}, {Chilingarian}, {Cikota}, {Colak}, {Colombo}, {Contreras}, {Cortina}, {Covino}, {D'Amico}, {D'Elia}, {Da Vela}, {Dazzi}, {De Angelis}, {De Lotto}, {Del Popolo}, {Delfino}, {Delgado}, {Delgado Mendez}, {Depaoli}, {Di Pierro}, {Di Venere}, {Do Souto Espi{\~n}eira}, {Prester}, {Donini}, {Dorner}, {Doro}, {Elsaesser}, {Fallah Ramazani}, {Fari{\~n}a Alonso}, {Fattorini}, {Fonseca}, {Font}, {Fruck}, {Fukami}, {Fukazawa}, {Garc{\'\i}a L{\'o}pez}, {Garczarczyk}, {Gasparyan}, {Gaug}, {Giglietto}, {Giordano}, {Gliwny}, {Godinovi{\'c}}, {Green}, {Green}, {Hadasch}, {Hahn}, {Hassan}, {Heckmann},
  {Herrera}, {Hoang}, {Hrupec}, {H{\"u}tten}, {Inada}, {Ishio}, {Iwamura}, {Jim{\'e}nez Mart{\'\i}nez}, {Jormanainen}, {Jouvin}, {Kerszberg}, {Kobayashi}, {Kubo}, {Kushida}, {Lamastra}, {Lelas}, {Leone}, {Lindfors}, {Linhoff}, {Lombardi}, {Longo}, {L{\'o}pez-Coto}, {L{\'o}pez-Moya}, {L{\'o}pez-Oramas}, {Loporchio}, {Machado de Oliveira Fraga}, {Maggio}, {Majumdar}, {Makariev}, {Mallamaci}, {Maneva}, {Manganaro}, {Mannheim}, {Maraschi}, {Mariotti}, {Mart{\'\i}nez}, {Mas Aguilar}, {Mazin}, {Menchiari}, {Mender}, {Mi{\'c}anovi{\'c}}, {Miceli}, {Miener}, {Miranda}, {Mirzoyan}, {Molina}, {Moralejo}, {Morcuende}, {Moreno}, {Moretti}, {Nakamori}, {Nava}, {Neustroev}, {Nievas Rosillo}, {Nigro}, {Nilsson}, {Nishijima}, {Noda}, {Nozaki}, {Ohtani}, {Oka}, {Otero-Santos}, {Paiano}, {Palatiello}, {Paneque}, {Paoletti}, {Paredes}, {Pavleti{\'c}}, {Pe{\~n}il}, {Persic}, {Pihet}, {Prada Moroni}, {Prandini}, {Priyadarshi}, {Puljak}, {Rhode}, {Rib{\'o}}, {Rico}, {Righi}, {Rugliancich}, {Sahakyan}, {Saito}, {Sakurai},
  {Satalecka}, {Saturni}, {Schleicher}, {Schmidt}, {Schweizer}, {Sitarek}, {{\v{S}}nidari{\'c}}, {Sobczynska}, {Spolon}, {Stamerra}, {Stri{\v{s}}kovi{\'c}}, {Strom}, {Strzys}, {Suda}, {Suri{\'c}}, {Takahashi}, {Takeishi}, {Tavecchio}, {Temnikov}, {Terzi{\'c}}, {Teshima}, {Tosti}, {Truzzi}, {Tutone}, {Ubach}, {van Scherpenberg}, {Vanzo}, {Vazquez Acosta}, {Ventura}, {Verguilov}, {Vigorito}, {Vitale}, {Vovk}, {Will}, {Wunderlich}, {Yamamoto}, {Zari{\'c}}, {Ambrosino}, {Cecconi}, {Catanzaro}, {Ferrara}, {Frasca}, {Munari}, {Giustolisi}, {Alonso-Santiago}, {Giarrusso}, {Munari}, \& {Valisa}}]{Acciari2022}
{Acciari}, V.~A., {Ansoldi}, S., {Antonelli}, L.~A., {et~al.} 2022, Nature Astronomy, 6, 689, \dodoi{10.1038/s41550-022-01640-z}

\bibitem[{{Acero, F.} {et~al.}(2010){Acero, F.}, {Aharonian, F.}, {Akhperjanian, A. G.}, {Anton, G.}, {Barres de Almeida, U.}, {Bazer-Bachi, A. R.}, {Becherini, Y.}, {Behera, B.}, {Beilicke, M.}, {Bernlöhr, K.}, {Bochow, A.}, {Boisson, C.}, {Bolmont, J.}, {Borrel, V.}, {Brucker, J.}, {Brun, F.}, {Brun, P.}, {Bühler, R.}, {Bulik, T.}, {Büsching, I.}, {Boutelier, T.}, {Chadwick, P. M.}, {Charbonnier, A.}, {Chaves, R. C. G.}, {Cheesebrough, A.}, {Conrad, J.}, {Chounet, L.-M.}, {Clapson, A. C.}, {Coignet, G.}, {Dalton, M.}, {Daniel, M. K.}, {Davids, I. D.}, {Degrange, B.}, {Deil, C.}, {Dickinson, H. J.}, {Djannati-Ataï, A.}, {Domainko, W.}, {Drury, L. O'C.}, {Dubois, F.}, {Dubus, G.}, {Dyks, J.}, {Dyrda, M.}, {Egberts, K.}, {Eger, P.}, {Espigat, P.}, {Fallon, L.}, {Farnier, C.}, {Fegan, S.}, {Feinstein, F.}, {Fiasson, A.}, {Förster, A.}, {Fontaine, G.}, {Füßling, M.}, {Gabici, S.}, {Gallant, Y. A.}, {Gérard, L.}, {Gerbig, D.}, {Giebels, B.}, {Glicenstein, J. F.}, {Glück, B.}, {Goret, P.}, {Göring, D.},
  {Hauser, D.}, {Hauser, M.}, {Heinz, S.}, {Heinzelmann, G.}, {Henri, G.}, {Hermann, G.}, {Hinton, J. A.}, {Hoffmann, A.}, {Hofmann, W.}, {Hofverberg, P.}, {Holleran, M.}, {Hoppe, S.}, {Horns, D.}, {Jacholkowska, A.}, {de Jager, O. C.}, {Jahn, C.}, {Jung, I.}, {Katarzyński, K.}, {Katz, U.}, {Kaufmann, S.}, {Kerschhaggl, M.}, {Khangulyan, D.}, {Khélifi, B.}, {Keogh, D.}, {Klochkov, D.}, {Kluźniak, W.}, {Kneiske, T.}, {Komin, Nu.}, {Kosack, K.}, {Kossakowski, R.}, {Lamanna, G.}, {Lemoine-Goumard, M.}, {Lenain, J.-P.}, {Lohse, T.}, {Marandon, V.}, {Marcowith, A.}, {Masbou, J.}, {Maurin, D.}, {McComb, T. J. L.}, {Medina, M. C.}, {Méhault, J.}, {Moderski, R.}, {Moulin, E.}, {Naumann-Godo, M.}, {de Naurois, M.}, {Nedbal, D.}, {Nekrassov, D.}, {Nicholas, B.}, {Niemiec, J.}, {Nolan, S. J.}, {Ohm, S.}, {Olive, J.-F.}, {de Oña Wilhelmi, E.}, {Orford, K. J.}, {Ostrowski, M.}, {Panter, M.}, {Paz Arribas, M.}, {Pedaletti, G.}, {Pelletier, G.}, {Petrucci, P.-O.}, {Pita, S.}, {Pühlhofer, G.}, {Punch, M.},
  {Quirrenbach, A.}, {Raubenheimer, B. C.}, {Raue, M.}, {Rayner, S. M.}, {Reimer, O.}, {Renaud, M.}, {de los Reyes, R.}, {Rieger, F.}, {Ripken, J.}, {Rob, L.}, {Rosier-Lees, S.}, {Rowell, G.}, {Rudak, B.}, {Rulten, C. B.}, {Ruppel, J.}, {Ryde, F.}, {Sahakian, V.}, {Santangelo, A.}, {Schlickeiser, R.}, {Schöck, F. M.}, {Schönwald, A.}, {Schwanke, U.}, {Schwarzburg, S.}, {Schwemmer, S.}, {Shalchi, A.}, {Sushch, I.}, {Sikora, M.}, {Skilton, J. L.}, {Sol, H.}, {Stawarz, Ł.}, {Steenkamp, R.}, {Stegmann, C.}, {Stinzing, F.}, {Superina, G.}, {Szostek, A.}, {Tam, P. H.}, {Tavernet, J.-P.}, {Terrier, R.}, {Tibolla, O.}, {Tluczykont, M.}, {van Eldik, C.}, {Vasileiadis, G.}, {Venter, C.}, {Venter, L.}, {Vialle, J. P.}, {Vincent, P.}, {Vink, J.}, {Vivier, M.}, {Völk, H. J.}, {Volpe, F.}, {Vorobiov, S.}, {Wagner, S. J.}, {Ward, M.}, {Zdziarski, A. A.}, {Zech, A.}, \& {HESS collaboration}}]{art:SN1006}
{Acero, F.}, {Aharonian, F.}, {Akhperjanian, A. G.}, {et~al.} 2010, A\&A, 516, A62, \dodoi{10.1051/0004-6361/200913916}

\bibitem[{Achterberg {et~al.}(2006)Achterberg, Ackermann, Adams, Ahrens, Andeen, Atlee, Baccus, Bahcall, Bai, Baret, Bartelt, Barwick, Bay, Beattie, Becka, Becker, Becker, Berghaus, Berley, Bernardini, Bertrand, Besson, Blaufuss, Boersma, Bohm, Böser, Botner, Bouchta, Braun, Burgess, Burgess, Castermans, Cherwinka, Chirkin, Clem, Cowen, D’Agostino, Davour, Day, {De Clercq}, Demirörs, Desiati, DeYoung, Diaz-Velez, Dreyer, Duvoort, Edwards, Ehrlich, Eisch, Elcheikh, Ellsworth, Evenson, Fadiran, Fazely, Feser, Filimonov, Fox, Gaisser, Gallagher, Ganugapati, Geenen, Gerhardt, Goldschmidt, Goodman, Gozzini, Greene, Grullon, Groß, Gunasingha, Gurtner, Hallgren, Halzen, Han, Hanson, Hardtke, Hardtke, Harenberg, Hart, Haugen, Hauschildt, Hays, Heise, Helbing, Hellwig, Herquet, Hill, Hodges, Hoffman, Hoshina, Hubert, Hughey, Hulth, Hultqvist, Hundertmark, Hülß, Ishihara, Jacobsen, Japaridze, Jones, Joseph, Kampert, Karle, Kawai, Kelley, Kestel, Kitamura, Klein, Klepser, Kohnen, Kolanoski, Köpke, Krasberg,
  Kuehn, Landsman, Laundrie, Leich, Liubarsky, Lundberg, Mackenzie, Madsen, Mase, Matis, McCauley, McParland, Meli, Messarius, Mészáros, Miyamoto, Mokhtarani, Montaruli, Morey, Morse, Movit, Münich, Muratas, Nahnhauer, Nam, Nießen, Nygren, Ögelman, Olbrechts, Olivas, Patton, Peña-Garay, {Pérez de los Heros}, Pettersen, Piegsa, Pieloth, Pohl, Porrata, Pretz, Price, Przybylski, Rawlins, Razzaque, Refflinghaus, Resconi, Rhode, Ribordy, Rizzo, Robbins, Rott, Rutledge, Sander, Sandstrom, Sarkar, Schlenstedt, Schneider, Seckel, Seo, Seunarine, Silvestri, Smith, Solarz, Song, Sopher, Spiczak, Spiering, Stamatikos, Stanev, Steffen, Stezelberger, Stokstad, Stoufer, Stoyanov, Strahler, Sulanke, Sullivan, Taboada, Tarasova, Tepe, Thollander, Tilav, Toale, Turčan, {van Eijndhoven}, Vandenbroucke, {Van Overloop}, Voigt, Wagner, Walck, Waldmann, Walter, Wang, Wendt, Whitney, Wiebusch, Wikström, Williams, Wischnewski, Wisniewski, Wissing, Woschnagg, Xu, Yodh, Yoshida, \& Zornoza}]{art:detIce}
Achterberg, A., Ackermann, M., Adams, J., {et~al.} 2006, Astroparticle Physics, 26, 155, \dodoi{https://doi.org/10.1016/j.astropartphys.2006.06.007}

\bibitem[{Ackermann {et~al.}(2013)Ackermann, Ajello, Allafort, Baldini, Ballet, Barbiellini, Baring, Bastieri, Bechtol, Bellazzini, Blandford, Bloom, Bonamente, Borgland, Bottacini, Brandt, Bregeon, Brigida, Bruel, \& Zimmer}]{art:lat_snr2}
Ackermann, M., Ajello, M., Allafort, A., {et~al.} 2013, Science (New York, N.Y.), 339, 807, \dodoi{10.1126/science.1231160}

\bibitem[{{Adams} {et~al.}(2021){Adams}, {Benbow}, {Brill}, {Buckley}, {Capasso}, {Chromey}, {Errando}, {Falcone}, {A Farrell}, {Feng}, {Finley}, {M Foote}, {Fortson}, {Furniss}, {Gent}, {Gillanders}, {Giuri}, {Gueta}, {Hanna}, {Hassan}, {Hervet}, {Holder}, {Hona}, {Humensky}, {Jin}, {Kaaret}, {Kertzman}, {Kieda}, {K Kleiner}, {Krennrich}, {Kumar}, {Lang}, {Lundy}, {Maier}, {E McGrath}, {Moriarty}, {Mukherjee}, {Nieto}, {Nievas-Rosillo}, {O'Brien}, {Ong}, {Otte}, {Park}, {Patel}, {Pfrang}, {Pichel}, {Pohl}, {Prado}, {Quinn}, {Ragan}, {Reynolds}, {Ribeiro}, {Roache}, {Rovero}, {Ryan}, {Santander}, {Schlenstedt}, {Sembroski}, {Shang}, {Tak}, {Vassiliev}, {Weinstein}, {Williams}, {J Williamson}, {J Williamson}, {Acciari}, {Ansoldi}, {Antonelli}, {Arbet Engels}, {Artero}, {Asano}, {Baack}, {Babi{\'c}}, {Baquero}, {Barres de Almeida}, {Barrio}, {Batkovi{\'c}}, {Becerra Gonz{\'a}lez}, {Bednarek}, {Bellizzi}, {Bernardini}, {Bernardos}, {Berti}, {Besenrieder}, {Bhattacharyya}, {Bigongiari}, {Biland}, {Blanch},
  {B{\"o}kenkamp}, {Bonnoli}, {Bo{\v{s}}njak}, {Busetto}, {Carosi}, {Ceribella}, {Cerruti}, {Chai}, {Chilingarian}, {Cikota}, {Colak}, {Colombo}, {Contreras}, {Cortina}, {Covino}, {D'Amico}, {D'Elia}, {Da Vela}, {Dazzi}, {De Angelis}, {De Lotto}, {Delfino}, {Delgado}, {Delgado Mendez}, {Depaoli}, {Di Pierro}, {Di Venere}, {Do Souto Espi{\~n}eira}, {Dominis Prester}, {Donini}, {Dorner}, {Doro}, {Elsaesser}, {Fallah Ramazani}, {Fattorini}, {Fonseca}, {Font}, {Fruck}, {Fukami}, {Fukazawa}, {Garc{\'\i}a L{\'o}pez}, {Garczarczyk}, {Gasparyan}, {Gaug}, {Giglietto}, {Giordano}, {Gliwny}, {Godinovi{\'c}}, {Green}, {Green}, {Hadasch}, {Hahn}, {Heckmann}, {Herrera}, {Hoang}, {Hrupec}, {H{\"u}tten}, {Inada}, {Ishio}, {Iwamura}, {Jim{\'e}nez Mart{\'\i}nez}, {Jormanainen}, {Jouvin}, {Karjalainen}, {Kerszberg}, {Kobayashi}, {Kubo}, {Kushida}, {Lamastra}, {Lelas}, {Leone}, {Lindfors}, {Linhoff}, {Lombardi}, {Longo}, {L{\'o}pez-Coto}, {L{\'o}pez-Moya}, {L{\'o}pez-Oramas}, {Loporchio}, {Machado de Oliveira Fraga}, {Maggio},
  {Majumdar}, {Makariev}, {Mallamaci}, {Maneva}, {Manganaro}, {Mannheim}, {Maraschi}, {Mariotti}, {Mart{\'\i}nez}, {Mazin}, {Menchiari}, {Mender}, {Mi{\'c}anovi{\'c}}, {Miceli}, {Miener}, {Miranda}, {Mirzoyan}, {Molina}, {Moralejo}, {Morcuende}, {Moreno}, {Moretti}, {Nakamori}, {Nava}, {Neustroev}, {Nigro}, {Nilsson}, {Nishijima}, {Noda}, {Nozaki}, {Ohtani}, {Oka}, {Otero-Santos}, {Paiano}, {Palatiello}, {Paneque}, {Paoletti}, {Paredes}, {Pavleti{\'c}}, {Pe{\~n}il}, {Persic}, {Pihet}, {Prada Moroni}, {Prandini}, {Priyadarshi}, {Puljak}, {Rhode}, {Rib{\'o}}, {Rico}, {Righi}, {Rugliancich}, {Saha}, {Sahakyan}, {Saito}, {Sakurai}, {Satalecka}, {Saturni}, {Schleicher}, {Schmidt}, {Schweizer}, {Sitarek}, {{\v{S}}nidari{\'c}}, {Sobczynska}, {Spolon}, {Stamerra}, {Stri{\v{s}}kovi{\'c}}, {Strom}, {Strzys}, {Suda}, {Suri{\'c}}, {Takahashi}, {Takeishi}, {Tavecchio}, {Temnikov}, {Terzi{\'c}}, {Teshima}, {Tosti}, {Truzzi}, {Tutone}, {Ubach}, {van Scherpenberg}, {Vanzo}, {Vazquez Acosta}, {Ventura}, {Verguilov},
  {Vigorito}, {Vitale}, {Vovk}, {Will}, {Wunderlich}, {Yamamoto}, {Zari{\'c}}, {Zari{\'c}}, {Abdalla}, {Aharonian}, {Ait Benkhali}, {Ang{\"u}ner}, {Arcaro}, {Ashkar}, {Backes}, {Barbosa Martins}, {Barnard}, {Batzofin}, {Becherini}, {Berge}, {Bernl{\"o}hr}, {Bi}, {B{\"o}ttcher}, {Boisson}, {Bolmont}, {de Bony de Lavergne}, {Breuhaus}, {Brose}, {Brun}, {Bulik}, {Caroff}, {Casanova}, {Chand}, {Chen}, {Cotter}, {Damascenev Mbarubucyeye}, {Devin}, {Djannati-Ata{\"\i}}, {Egberts}, {Ernenwein}, {Fegan}, {Fiasson}, {Fichet de Clairfontaine}, {Fontaine}, {F{\"u}{\ss}ling}, {Funk}, {Gabici}, {Giavitto}, {Glawion}, {Glicenstein}, {Grondin}, {Hinton}, {Hofmann}, {Holch}, {Holler}, {Horns}, {Huang}, {Jamrozy}, {Jankowsky}, {Joshi}, {Jung-Richardt}, {Kasai}, {Katarzy{\'n}ski}, {Kh{\'e}lifi}, {Komin}, {Kosack}, {Kostunin}, {Le Stum}, {Lemi{\`e}re}, {Lenain}, {Leuschner}, {Levy}, {Lohse}, {Luashvili}, {Lypova}, {Mackey}, {Majumdar}, {Malyshev}, {Marandon}, {Marchegiani}, {Marcowith}, {Mart{\'\i}-Devesa}, {Marx}, {Maurin},
  {Meintjes}, {Mitchell}, {Moderski}, {Mohrmann}, {Montanari}, {Moulin}, {Muller}, {Murach}, {de Naurois}, {Nayerhoda}, {Niemiec}, {Priyana Noel}, {O'Brien}, {Ohm}, {Olivera-Nieto}, {de Ona Wilhelmi}, {Ostrowski}, {Panny}, {Panter}, {Parsons}, {Peron}, {Poireau}, {Prokhorov}, {Prokoph}, {P{\"u}hlhofer}, {Punch}, {Quirrenbach}, {Reichherzer}, {Reimer}, {Reimer}, {Renaud}, {Rieger}, {Romoli}, {Rowell}, {Rudak}, {Rueda Ricarte}, {Ruiz-Velasco}, {Sahakian}, {Sailer}, {Salzmann}, {Sanchez}, {Santangelo}, {Sasaki}, {Schutte}, {Schwanke}, {Sch{\"u}ssler}, {Senniappan}, {Shapopi}, {Simoni}, {Sol}, {Specovius}, {Spencer}, {Steenkamp}, {Steinmassl}, {Sun}, {Takahashi}, {Tanaka}, {Terrier}, {Tsuji}, {Uchiyama}, {van Eldik}, {van Soelen}, {Veh}, {Venter}, {Vink}, {Wagner}, {White}, {Wierzcholska}, {Wun Wong}, {Zacharias}, {Zargaryan}, {Zdziarski}, {Zech}, {Zhu}, {Zouari}, {{\.Z}ywucka}, {{\.Z}ywucka}, {Moritani}, \& {Torres}}]{HESSJ0632}
{Adams}, C.~B., {Benbow}, W., {Brill}, A., {et~al.} 2021, \apj, 923, 241, \dodoi{10.3847/1538-4357/ac29b7}

\bibitem[{Adrián-Martínez {et~al.}(2016)Adrián-Martínez, Ageron, Aharonian, Aiello, Albert, Ameli, Anassontzis, André, Androulakis, Anghinolfi, Anton, Ardid, Avgitas, Barbarino, Barbarito, Baret, Barrios-Martí, Belhorma, Belias, \& Zuñiga}]{art:loi}
Adrián-Martínez, S., Ageron, M., Aharonian, F., {et~al.} 2016, Journal of Physics G: Nuclear and Particle Physics, 43, \dodoi{10.1088/0954-3899/43/8/084001}

\bibitem[{Ageron {et~al.}(2011)Ageron, Aguilar, {Al Samarai}, Albert, Ameli, André, Anghinolfi, Anton, Anvar, Ardid, Arnaud, Aslanides, {Assis Jesus}, Astraatmadja, Aubert, Auer, Barbarito, Baret, Basa, Bazzotti, Becherini, Beltramelli, Bersani, Bertin, Beurthey, Biagi, Bigongiari, Billault, Blaes, Bogazzi, {de Botton}, Bou-Cabo, Boudahef, Bouwhuis, Brown, Brunner, Busto, Caillat, Calzas, Camarena, Capone, Caponetto, Cârloganu, Carminati, Carmona, Carr, Carton, Cassano, Castorina, Cecchini, Ceres, Chaleil, Charvis, Chauchot, Chiarusi, Circella, Compère, Coniglione, Coppolani, Cosquer, Costantini, Cottini, Coyle, Cuneo, Curtil, D'Amato, Damy, {van Dantzig}, {De Bonis}, Decock, Decowski, Dekeyser, Delagnes, Desages-Ardellier, Deschamps, Destelle, {Di Maria}, Dinkespiler, Distefano, Dominique, Donzaud, Dornic, Dorosti, Drogou, Drouhin, Druillole, Durand, Durand, Eberl, Emanuele, Engelen, Ernenwein, Escoffier, Falchini, Favard, Fehr, Feinstein, Ferri, Ferry, Fiorello, Flaminio, Folger, Fritsch, Fuda, Galatá,
  Galeotti, Gay, Gensolen, Giacomelli, Gojak, Gómez-González, Goret, Graf, Guillard, Halladjian, Hallewell, {van Haren}, Hartmann, Heijboer, Heine, Hello, Henry, Hernández-Rey, Herold, Hößl, Hogenbirk, Hsu, Hubbard, Jaquet, Jaspers, {de Jong}, Jourde, Kadler, Kalantar-Nayestanaki, Kalekin, Kappes, Karg, Karkar, Karolak, Katz, Keller, Kestener, Kok, Kok, Kooijman, Kopper, Kouchner, Kretschmer, Kruijer, Kuch, Kulikovskiy, Lachartre, Lafoux, Lagier, Lahmann, Lahonde-Hamdoun, Lamare, Lambard, Languillat, Larosa, Lavalle, {Le Guen}, {Le Provost}, LeVanSuu, Lefèvre, Legou, Lelaizant, Lévéque, Lim, {Lo Presti}, Loehner, Loucatos, Louis, Lucarelli, Lyashuk, Magnier, Mangano, Marcel, Marcelin, Margiotta, Martinez-Mora, Masullo, Mazéas, Mazure, Meli, Melissas, Migneco, Mongelli, Montaruli, Morganti, Moscoso, Motz, Musumeci, Naumann, Naumann-Godo, Neff, Niess, Nooren, Oberski, Olivetto, Palanque-Delabrouille, Palioselitis, Papaleo, Păvălaş, Payet, Payre, Peek, Petrovic, Piattelli, Picot-Clemente, Picq,
  Piret, Poinsignon, Popa, Pradier, Presani, Prono, Racca, Raia, {van Randwijk}, Real, Reed, Réthoré, Rewiersma, Riccobene, Richardt, Richter, Ricol, Rigaud, Roca, Roensch, Rolin, Rostovtsev, Rottura, Roux, Rujoiu, Ruppi, Russo, Salesa, Salomon, Sapienza, Schmitt, Schöck, Schuller, Schüssler, Sciliberto, Shanidze, Shirokov, Simeone, Sottoriva, Spies, Spona, Spurio, Steijger, Stolarczyk, Streeb, Sulak, Taiuti, Tamburini, Tao, Tasca, Terreni, Tezier, Toscano, Urbano, Valdy, Vallage, {Van Elewyck}, Vannoni, Vecchi, Venekamp, Verlaat, Vernin, Virique, {de Vries}, {van Wijk}, Wijnker, Wobbe, {de Wolf}, Yakovenko, Yepes, Zaborov, Zaccone, Zornoza, \& Zúñiga}]{detAN}
Ageron, M., Aguilar, J., {Al Samarai}, I., {et~al.} 2011, Nuclear Instruments and Methods in Physics Research Section A: Accelerators, Spectrometers, Detectors and Associated Equipment, 656, 11, \dodoi{https://doi.org/10.1016/j.nima.2011.06.103}

\bibitem[{{Aharonian} {et~al.}(2004){Aharonian}, {Akhperjanian}, {Beilicke}, {Bernl{\"o}hr}, {B{\"o}rst}, {Bojahr}, {Bolz}, {Coarasa}, {Contreras}, {Cortina}, {Denninghoff}, {Fonseca}, {Girma}, {G{\"o}tting}, {Heinzelmann}, {Hermann}, {Heusler}, {Hofmann}, {Horns}, {Jung}, {Kankanyan}, {Kestel}, {Kohnle}, {Konopelko}, {Kranich}, {Lampeitl}, {Lopez}, {Lorenz}, {Lucarelli}, {Mang}, {Mazin}, {Meyer}, {Mirzoyan}, {Moralejo}, {O{\~n}a-Wilhelmi}, {Panter}, {Plyasheshnikov}, {P{\"u}hlhofer}, {de los Reyes}, {Rhode}, {Ripken}, {Rowell}, {Sahakian}, {Samorski}, {Schilling}, {Siems}, {Sobzynska}, {Stamm}, {Tluczykont}, {Vitale}, {V{\"o}lk}, {Wiedner}, \& {Wittek}}]{Crab}
{Aharonian}, F., {Akhperjanian}, A., {Beilicke}, M., {et~al.} 2004, \apj, 614, 897, \dodoi{10.1086/423931}

\bibitem[{{Ahnen} {et~al.}(2016){Ahnen}, {Ansoldi}, {Antonelli}, {Antoranz}, {Babic}, {Banerjee}, {Bangale}, {Barres de Almeida}, {Barrio}, {Becerra Gonz{\'a}lez}, {Bednarek}, {Bernardini}, {Biasuzzi}, {Biland}, {Blanch}, {Bonnefoy}, {Bonnoli}, {Borracci}, {Bretz}, {Buson}, {Carosi}, {Chatterjee}, {Clavero}, {Colin}, {Colombo}, {Contreras}, {Cortina}, {Covino}, {Da Vela}, {Dazzi}, {De Angelis}, {De Lotto}, {de O{\~n}a Wilhelmi}, {Delgado Mendez}, {Di Pierro}, {Dom{\'\i}nguez}, {Dominis Prester}, {Dorner}, {Doro}, {Einecke}, {Eisenacher Glawion}, {Elsaesser}, {Fern{\'a}ndez-Barral}, {Fidalgo}, {Fonseca}, {Font}, {Frantzen}, {Fruck}, {Galindo}, {Garc{\'\i}a L{\'o}pez}, {Garczarczyk}, {Garrido Terrats}, {Gaug}, {Giammaria}, {Godinovi{\'c}}, {Gonz{\'a}lez Mu{\~n}oz}, {Gora}, {Guberman}, {Hadasch}, {Hahn}, {Hanabata}, {Hayashida}, {Herrera}, {Hose}, {Hrupec}, {Hughes}, {Idec}, {Kodani}, {Konno}, {Kubo}, {Kushida}, {La Barbera}, {Lelas}, {Lindfors}, {Lombardi}, {Longo}, {L{\'o}pez}, {L{\'o}pez-Coto},
  {L{\'o}pez-Oramas}, {Majumdar}, {Makariev}, {Mallot}, {Maneva}, {Manganaro}, {Mannheim}, {Maraschi}, {Marcote}, {Mariotti}, {Mart{\'\i}nez}, {Mazin}, {Menzel}, {Miranda}, {Mirzoyan}, {Moralejo}, {Moretti}, {Nakajima}, {Neustroev}, {Niedzwiecki}, {Nievas Rosillo}, {Nilsson}, {Nishijima}, {Noda}, {Orito}, {Overkemping}, {Paiano}, {Palacio}, {Palatiello}, {Paneque}, {Paoletti}, {Paredes}, {Paredes-Fortuny}, {Pedaletti}, {Persic}, {Poutanen}, {Prada Moroni}, {Prandini}, {Puljak}, {Rhode}, {Rib{\'o}}, {Rico}, {Rodriguez Garcia}, {Saito}, {Satalecka}, {Schultz}, {Schweizer}, {Shore}, {Sillanp{\"a}{\"a}}, {Sitarek}, {Snidaric}, {Sobczynska}, {Stamerra}, {Steinbring}, {Strzys}, {Takalo}, {Takami}, {Tavecchio}, {Temnikov}, {Terzi{\'c}}, {Tescaro}, {Teshima}, {Thaele}, {Torres}, {Toyama}, {Treves}, {Verguilov}, {Vovk}, {Ward}, {Will}, {Wu}, {Zanin}, {MAGIC Collaboration}, {Casares}, \& {Herrero}}]{art:LS+61_tev}
{Ahnen}, M.~L., {Ansoldi}, S., {Antonelli}, L.~A., {et~al.} 2016, \aap, 591, A76, \dodoi{10.1051/0004-6361/201527964}

\bibitem[{{Aiello} {et~al.}(2019){Aiello}, {Akrame}, {Ameli}, {Anassontzis}, {Andre}, {Androulakis}, {Anghinolfi}, {Anton}, {Ardid}, {Aublin}, {Avgitas}, {Bagatelas}, {Barbarino}, {Baret}, {Barrios-Mart{\'\i}}, {Belias}, {Berbee}, {van den Berg}, {Bertin}, {Biagi}, {Biagioni}, {Biernoth}, {Boumaaza}, {Bourret}, {Bouta}, {Bouwhuis}, {Bozza}, {Br{\^a}nza{\c{s}}}, {Bruchner}, {Bruijn}, {Brunner}, {Buis}, {Buompane}, {Busto}, {Calvo}, {Capone}, {Celli}, {Chabab}, {Chau}, {Cherubini}, {Chiarella}, {Chiarusi}, {Circella}, {Cocimano}, {Coelho}, {Coleiro}, {Molla}, {Coniglione}, {Coyle}, {Creusot}, {Cuttone}, {D'Onofrio}, {Dallier}, {De Sio}, {Di Palma}, {D{\'\i}az}, {Diego-Tortosa}, {Distefano}, {Domi}, {Don{\`a}}, {Donzaud}, {Dornic}, {D{\"o}rr}, {Durocher}, {Eberl}, {van Eijk}, {El Bojaddaini}, {Eljarrari}, {Elsaesser}, {Enzenh{\"o}fer}, {Fermani}, {Ferrara}, {Filipovi{\'c}}, {Fusco}, {Gal}, {Garcia}, {Garufi}, {Gialanella}, {Giorgio}, {Giuliante}, {Gozzini}, {Gracia}, {Graf}, {Grasso}, {Gr{\'e}goire}, {Grella},
  {Hallmann}, {Hamdaoui}, {van Haren}, {Heid}, {Heijboer}, {Hekalo}, {Hern{\'a}ndez-Rey}, {Hofest{\"a}dt}, {Illuminati}, {James}, {Jongen}, {de Jong}, {de Jong}, {Kadler}, {Kalaczy{\'n}ski}, {Kalekin}, {Katz}, {Khan Chowdhury}, {Kie{\ss}ling}, {Koffeman}, {Kooijman}, {Kouchner}, {Kreter}, {Kulikovskiy}, {Kunhikannan Kannichankandy}, {Lahmann}, {Larosa}, {Le Breton}, {Leone}, {Leonora}, {Levi}, {Lincetto}, {Lonardo}, {Longhitano}, {Lopez Coto}, {Lotze}, {Maderer}, {Maggi}, {Ma{\'n}czak}, {Mannheim}, {Margiotta}, {Marinelli}, {Markou}, {Martin}, {Mart{\'\i}nez-Mora}, {Martini}, {Marzaioli}, {Mele}, {Melis}, {Migliozzi}, {Migneco}, {Mijakowski}, {Miranda}, {Mollo}, {Morganti}, {Moser}, {Moussa}, {Muller}, {Musumeci}, {Nauta}, {Navas}, {Nicolau}, {Nielsen}, {{\'O} Fearraigh}, {Organokov}, {Orlando}, {Ottonello}, {Panagopoulos}, {Papalashvili}, {Papaleo}, {P{\u{a}}v{\u{a}}la{\c{s}}}, {Pellegrino}, {Perrin-Terrin}, {Piattelli}, {Pikounis}, {Pisanti}, {Poir{\'e}}, {Polydefki}, {Popa}, {Post}, {Pradier},
  {P{\"u}hlhofer}, {Pulvirenti}, {Quinn}, {Raffaelli}, {Randazzo}, {Razzaque}, {Real}, {Resvanis}, {Reubelt}, {Riccobene}, {Richer}, {Rigalleau}, {Rovelli}, {Saffer}, {Salvadori}, {Samtleben}, {S{\'a}nchez Losa}, {Sanguineti}, {Santangelo}, {Santonocito}, {Sapienza}, {Schumann}, {Sciacca}, {Seneca}, {Sgura}, {Shanidze}, {Sharma}, {Simeone}, {Sinopoulou}, {Spisso}, {Spurio}, {Stavropoulos}, {Steijger}, {Stellacci}, {Strandberg}, {Stransky}, {St{\"u}ven}, {Taiuti}, {Tatone}, {Tayalati}, {Tenllado}, {Thakore}, {Trovato}, {Tzamariudaki}, {Tzanetatos}, {Van Elewyck}, {Versari}, {Viola}, {Vivolo}, {Wilms}, {de Wolf}, {Zaborov}, {Zornoza}, {Z{\'u}{\~n}iga}, \& {KM3NeT Collaboration}}]{art:km3_visibility}
{Aiello}, S., {Akrame}, S.~E., {Ameli}, F., {et~al.} 2019, Astroparticle Physics, 111, 100, \dodoi{10.1016/j.astropartphys.2019.04.002}

\bibitem[{Albert {et~al.}(2017)Albert, Andr\'e, Anghinolfi, Anton, Ardid, Aubert, Avgitas, Baret, Barrios-Mart\'{\i}, Basa, Belhorma, Bertin, Biagi, Bormuth, Bourret, Bouwhuis, Bruijn, Brunner, Busto, Capone, Caramete, Carr, Celli, Cherkaoui El~Moursli, Chiarusi, Circella, Coelho, Coleiro, Coniglione, Costantini, Coyle, Creusot, D\'{\i}az, Deschamps, De~Bonis, Distefano, Di~Palma, Domi, Donzaud, Dornic, Drouhin, Eberl, El~Bojaddaini, El~Khayati, Els\"asser, Enzenh\"ofer, Ettahiri, Fassi, Felis, Fusco, Galat\`a, Gay, Giordano, Glotin, Gr\'egoire, Gracia~Ruiz, Graf, Hallmann, van Haren, Heijboer, Hello, Hern\'andez-Rey, H\"o\ss{}l, Hofest\"adt, Hugon, Illuminati, James, de~Jong, Jongen, Kadler, Kalekin, Katz, Kie\ss{}ling, Kouchner, Kreter, Kreykenbohm, Kulikovskiy, Lachaud, Lahmann, Lef\`evre, Leonora, Lotze, Loucatos, Marcelin, Margiotta, Marinelli, Mart\'{\i}nez-Mora, Mele, Melis, Michael, Migliozzi, Moussa, Navas, Nezri, Organokov, P\ifmmode \u{a}\else \u{a}\fi{}v\ifmmode \u{a}\else
  \u{a}\fi{}la\ifmmode~\mbox{\c{s}}\else \c{s}\fi{}, Pellegrino, Perrina, Piattelli, Popa, Pradier, Quinn, Racca, Riccobene, S\'anchez-Losa, Salda\~na, Salvadori, Samtleben, Sanguineti, Sapienza, Sch\"ussler, Sieger, Spurio, Stolarczyk, Taiuti, Tayalati, Trovato, Turpin, T\"onnis, Vallage, Van~Elewyck, Versari, Vivolo, Vizzoca, Wilms, Zornoza, Z\'u\~niga, Gaggero, \& Grasso}]{antares2017}
Albert, A., Andr\'e, M., Anghinolfi, M., {et~al.} 2017, Phys. Rev. D, 96, 062001, \dodoi{10.1103/PhysRevD.96.062001}

\bibitem[{Albert {et~al.}(2018)Albert, André, Anghinolfi, Ardid, Aubert, Aublin, Avgitas, Baret, Barrios-Martí, Basa, Belhorma, Bertin, Biagi, Bormuth, Boumaaza, Bourret, Bouwhuis, Brânzaş, Bruijn, Brunner, Busto, Capone, Caramete, Carr, Celli, Chabab, Moursli, Chiarusi, Circella, Coelho, Coleiro, Colomer, Coniglione, Costantini, Coyle, Creusot, Díaz, Deschamps, Distefano, Palma, Domi, Donzaud, Dornic, Drouhin, Eberl, Bojaddaini, Khayati, Elsässer, Enzenhöfer, Ettahiri, Fassi, Felis, Fermani, Ferrara, Fusco, Gay, Glotin, Grégoire, Ruiz, Graf, Hallmann, van Haren, Heijboer, Hello, Hernández-Rey, Hößl, Hofestädt, Illuminati, James, de~Jong, Jongen, Kadler, Kalekin, Katz, Khan-Chowdhury, Kouchner, Kreter, Kreykenbohm, Kulikovskiy, Lachaud, Lahmann, Lefèvre, Leonora, Levi, Lotze, Loucatos, Marcelin, Margiotta, Marinelli, Martínez-Mora, Mele, Melis, Migliozzi, Moussa, Navas, Nezri, Nuñez, Organokov, Păvălaş, Pellegrino, Piattelli, Popa, Pradier, Quinn, Racca, Randazzo, Riccobene, Sánchez-Losa,
  Saldaña, Salvadori, Samtleben, Sanguineti, Sapienza, Schüssler, Spurio, Stolarczyk, Taiuti, Tayalati, Trovato, Vallage, Elewyck, Versari, Vivolo, Wilms, Zaborov, Zornoza, Zúñiga, ANTARES Collaboration, Aartsen, Ackermann, Adams, Aguilar, Ahlers, Ahrens, Samarai, Altmann, Andeen, Anderson, Ansseau, Anton, Argüelles, Auffenberg, Axani, Backes, Bagherpour, Bai, Barbano, Barron, Barwick, Baum, Bay, Beatty, Tjus, Becker, BenZvi, Berley, Bernardini, Besson, Binder, Bindig, Blaufuss, Blot, Bohm, Börner, Bos, Böser, Botner, Bourbeau, Bourbeau, Bradascio, Braun, Brenzke, Bretz, Bron, Brostean-Kaiser, Burgman, Busse, Carver, Cheung, Chirkin, Christov, Clark, Classen, Collin, Conrad, Coppin, Correa, Cowen, Cross, Dave, Day, de~André, Clercq, DeLaunay, Dembinski, Deoskar, Ridder, Desiati, de~Vries, de~Wasseige, de~With, DeYoung, Díaz-Vélez, di~Lorenzo, Dujmovic, Dumm, Dunkman, Dvorak, Eberhardt, Ehrhardt, Eichmann, Eller, Evenson, Fahey, Fazely, Felde, Filimonov, Finley, Franckowiak, Friedman, Fritz,
  Gaisser, Gallagher, Ganster, Gerhardt, Ghorbani, Giang, Glauch, Glüsenkamp, Goldschmidt, Gonzalez, Grant, Griffith, Haack, Hallgren, Halve, Halzen, Hanson, Hebecker, Heereman, Helbing, Hellauer, Hickford, Hignight, Hill, Hoffman, Hoffmann, Hoinka, Hokanson-Fasig, Hoshina, Huang, Huber, Hultqvist, Hünnefeld, Hussain, In, Iovine, Ishihara, Jacobi, Japaridze, Jeong, Jero, Jones, Kalaczynski, Kang, Kappes, Kappesser, Karg, Karle, Katz, Kauer, Keivani, Kelley, Kheirandish, Kim, Kintscher, Kiryluk, Kittler, Klein, Koirala, Kolanoski, Köpke, Kopper, Kopper, Koschinsky, Koskinen, Kowalski, Krings, Kroll, Krückl, Kunwar, Kurahashi, Kyriacou, Labare, Lanfranchi, Larson, Lauber, Leonard, Leuermann, Liu, Lohfink, Mariscal, Lu, Lünemann, Luszczak, Madsen, Maggi, Mahn, Makino, Mancina, Maruyama, Mase, Maunu, Meagher, Medici, Meier, Menne, Merino, Meures, Miarecki, Micallef, Momenté, Montaruli, Moore, Moulai, Nagai, Nahnhauer, Nakarmi, Naumann, Neer, Niederhausen, Nowicki, Nygren, Pollmann, Olivas, O’Murchadha,
  O’Sullivan, Palczewski, Pandya, Pankova, Peiffer, Pepper, de~los Heros, Pieloth, Pinat, Pizzuto, Plum, Price, Przybylski, Raab, Rameez, Rauch, Rawlins, Rea, Reimann, Relethford, Resconi, Rhode, Richman, Robertson, Rongen, Rott, Ruhe, Ryckbosch, Rysewyk, Safa, Herrera, Sandrock, Sandroos, Santander, Sarkar, Sarkar, Satalecka, Schaufel, Schlunder, Schmidt, Schneider, Schöneberg, Schumacher, Sclafani, Seckel, Seunarine, Soedingrekso, Soldin, Song, Spiczak, Spiering, Stachurska, Stamatikos, Stanev, Stasik, Stein, Stettner, Steuer, Stezelberger, Stokstad, Stößl, Strotjohann, Stuttard, Sullivan, Sutherland, Taboada, Tenholt, Ter-Antonyan, Terliuk, Tilav, Toale, Tobin, Tönnis, Toscano, Tosi, Tselengidou, Tung, Turcati, Turley, Ty, Unger, Elorrieta, Usner, Vandenbroucke, Driessche, van Eijk, van Eijndhoven, Vanheule, van Santen, Vraeghe, Walck, Wallace, Wallraff, Wandler, Wandkowsky, Watson, Waza, Weaver, Weiss, Wendt, Werthebach, Westerhoff, Whelan, Whitehorn, Wiebe, Wiebusch, Wille, Williams, Wills, Wolf,
  Wood, Wood, Woolsey, Woschnagg, Wrede, Xu, Xu, Xu, Yanez, Yodh, Yoshida, Yuan, IceCube Collaboration, Gaggero, \& Grasso}]{albert2018}
Albert, A., André, M., Anghinolfi, M., {et~al.} 2018, The Astrophysical Journal Letters, 868, L20, \dodoi{10.3847/2041-8213/aaeecf}

\bibitem[{{Albert} {et~al.}(2023){Albert}, {Alves}, {Andr{\'e}}, {Ardid}, {Ardid}, {Aubert}, {Aublin}, {Baret}, {Basa}, {Becherini}, {Belhorma}, {Bendahman}, {Benfenati}, {Bertin}, {Biagi}, {Bissinger}, {Boumaaza}, {Bouta}, {Bouwhuis}, {Br{\^a}nza{\c{s}}}, {Bruijn}, {Brunner}, {Busto}, {Caiffi}, {Calvo}, {Campion}, {Capone}, {Caramete}, {Carenini}, {Carr}, {Carretero}, {Celli}, {Cerisy}, {Chabab}, {Chau}, {Cherkaoui El Moursli}, {Chiarusi}, {Circella}, {Coelho}, {Coleiro}, {Coniglione}, {Coyle}, {Creusot}, {D{\'\i}az}, {de Martino}, {Distefano}, {di Palma}, {Domi}, {Donzaud}, {Dornic}, {Drouhin}, {Eberl}, {van Eeden}, {van Eijk}, {El Hedri}, {El Khayati}, {Enzenh{\"o}fer}, {Fasano}, {Fermani}, {Ferrara}, {Filippini}, {Fusco}, {Gagliardini}, {Garc{\'\i}a}, {Gatius Oliver}, {Gay}, {Gei{\ss}elbrecht}, {Glotin}, {Gozzini}, {Gracia Ruiz}, {Graf}, {Guidi}, {Haegel}, {Hallmann}, {van Haren}, {Heijboer}, {Hello}, {Hern{\'a}ndez-Rey}, {H{\"o}{\ss}l}, {Hofest{\"a}dt}, {Huang}, {Illuminati}, {James}, {Jisse-Jung}, {de
  Jong}, {de Jong}, {Kadler}, {Kalekin}, {Katz}, {Kouchner}, {Kreykenbohm}, {Kulikovskiy}, {Lahmann}, {Lamoureux}, {Lazo}, {Lef{\`e}vre}, {Leonora}, {Levi}, {Le Stum}, {Lopez-Coto}, {Loucatos}, {Maderer}, {Manczak}, {Marcelin}, {Margiotta}, {Marinelli}, {Mart{\'\i}nez-Mora}, {Migliozzi}, {Moussa}, {Muller}, {Nauta}, {Navas}, {Neronov}, {Nezri}, {{\'O} Fearraigh}, {P{\u{a}}un}, {P{\u{a}}v{\u{a}}la{\c{s}}}, {Perrin-Terrin}, {Pestel}, {Piattelli}, {Poir{\`e}}, {Popa}, {Pradier}, {Randazzo}, {Real}, {Reck}, {Riccobene}, {Romanov}, {S{\'a}nchez-Losa}, {Saina}, {Salesa Greus}, {Samtleben}, {Sanguineti}, {Sapienza}, {Savchenko}, {Schnabel}, {Schumann}, {Sch{\"u}ssler}, {Seneca}, {Spurio}, {Stolarczyk}, {Taiuti}, {Tayalati}, {Tingay}, {Vallage}, {Vannoye}, {van Elewyck}, {Viola}, {Vivolo}, {Wilms}, {Zavatarelli}, {Zegarelli}, {Zornoza}, {Z{\'u}{\~n}iga}, \& {ANTARES Collaboration}}]{antares2023}
{Albert}, A., {Alves}, S., {Andr{\'e}}, M., {et~al.} 2023, Physics Letters B, 841, 137951, \dodoi{10.1016/j.physletb.2023.137951}

\bibitem[{{Albert} {et~al.}(2007{\natexlab{a}}){Albert}, {Aliu}, {Anderhub}, {Antoranz}, {Armada}, {Baixeras}, {Barrio}, {Bartko}, {Bastieri}, {Becker}, {Bednarek}, {Berger}, {Bigongiari}, {Biland}, {Bock}, {Bordas}, {Bosch-Ramon}, {Bretz}, {Britvitch}, {Camara}, {Carmona}, {Chilingarian}, {Coarasa}, {Commichau}, {Contreras}, {Cortina}, {Costado}, {Curtef}, {Danielyan}, {Dazzi}, {De Angelis}, {Delgado}, {de los Reyes}, {De Lotto}, {Domingo-Santamar{\'\i}a}, {Dorner}, {Doro}, {Errando}, {Fagiolini}, {Ferenc}, {Fern{\'a}ndez}, {Firpo}, {Flix}, {Fonseca}, {Font}, {Fuchs}, {Galante}, {Garc{\'\i}a-L{\'o}pez}, {Garczarczyk}, {Gaug}, {Giller}, {Goebel}, {Hakobyan}, {Hayashida}, {Hengstebeck}, {Herrero}, {H{\"o}hne}, {Hose}, {Hsu}, {Jacon}, {Jogler}, {Kosyra}, {Kranich}, {Kritzer}, {Laille}, {Lindfors}, {Lombardi}, {Longo}, {L{\'o}pez}, {L{\'o}pez}, {Lorenz}, {Majumdar}, {Maneva}, {Mannheim}, {Mansutti}, {Mariotti}, {Mart{\'\i}nez}, {Mazin}, {Merck}, {Meucci}, {Meyer}, {Miranda}, {Mirzoyan}, {Mizobuchi}, {Moralejo},
  {Nieto}, {Nilsson}, {Ninkovic}, {O{\~n}a-Wilhelmi}, {Otte}, {Oya}, {Paneque}, {Panniello}, {Paoletti}, {Paredes}, {Pasanen}, {Pascoli}, {Pauss}, {Pegna}, {Persic}, {Peruzzo}, {Piccioli}, {Prandini}, {Puchades}, {Raymers}, {Rhode}, {Rib{\'o}}, {Rico}, {Rissi}, {Robert}, {R{\"u}gamer}, {Saggion}, {Saito}, {S{\'a}nchez}, {Sartori}, {Scalzotto}, {Scapin}, {Schmitt}, {Schweizer}, {Shayduk}, {Shinozaki}, {Shore}, {Sidro}, {Sillanp{\"a}{\"a}}, {Sobczynska}, {Stamerra}, {Stark}, {Takalo}, {Temnikov}, {Tescaro}, {Teshima}, {Torres}, {Turini}, {Vankov}, {Vitale}, {Wagner}, {Wibig}, {Wittek}, {Zandanel}, {Zanin}, \& {Zapatero}}]{IC443}
{Albert}, J., {Aliu}, E., {Anderhub}, H., {et~al.} 2007{\natexlab{a}}, ApJL, 664, L87, \dodoi{10.1086/520957}

\bibitem[{{Albert} {et~al.}(2007{\natexlab{b}}){Albert}, {Aliu}, {Anderhub}, {Antoranz}, {Armada}, {Baixeras}, {Barrio}, {Bartko}, {Bastieri}, {Becker}, {Bednarek}, {Berger}, {Bigongiari}, {Biland}, {Bock}, {Bordas}, {Bosch-Ramon}, {Bretz}, {Britvitch}, {Camara}, {Carmona}, {Chilingarian}, {Coarasa}, {Commichau}, {Contreras}, {Cortina}, {Costado}, {Curtef}, {Danielyan}, {Dazzi}, {De Angelis}, {Delgado}, {de los Reyes}, {De Lotto}, {Domingo-Santamar{\'\i}a}, {Dorner}, {Doro}, {Errando}, {Fagiolini}, {Ferenc}, {Fern{\'a}ndez}, {Firpo}, {Flix}, {Fonseca}, {Font}, {Fuchs}, {Galante}, {Garc{\'\i}a-L{\'o}pez}, {Garczarczyk}, {Gaug}, {Giller}, {Goebel}, {Hakobyan}, {Hayashida}, {Hengstebeck}, {Herrero}, {H{\"o}hne}, {Hose}, {Hsu}, {Jacon}, {Jogler}, {Kosyra}, {Kranich}, {Kritzer}, {Laille}, {Lindfors}, {Lombardi}, {Longo}, {L{\'o}pez}, {L{\'o}pez}, {Lorenz}, {Majumdar}, {Maneva}, {Mannheim}, {Mansutti}, {Mariotti}, {Mart{\'\i}nez}, {Mazin}, {Merck}, {Meucci}, {Meyer}, {Miranda}, {Mirzoyan}, {Mizobuchi}, {Moralejo},
  {Nieto}, {Nilsson}, {Ninkovic}, {O{\~n}a-Wilhelmi}, {Otte}, {Oya}, {Panniello}, {Paoletti}, {Paredes}, {Pasanen}, {Pascoli}, {Pauss}, {Pegna}, {Persic}, {Peruzzo}, {Piccioli}, {Prandini}, {Puchades}, {Raymers}, {Rhode}, {Rib{\'o}}, {Rico}, {Rissi}, {Robert}, {R{\"u}gamer}, {Saggion}, {Saito}, {S{\'a}nchez}, {Sartori}, {Scalzotto}, {Scapin}, {Schmitt}, {Schweizer}, {Shayduk}, {Shinozaki}, {Shore}, {Sidro}, {Sillanp{\"a}{\"a}}, {Sobczynska}, {Stamerra}, {Stark}, {Takalo}, {Temnikov}, {Tescaro}, {Teshima}, {Torres}, {Turini}, {Vankov}, {Vitale}, {Wagner}, {Wibig}, {Wittek}, {Zandanel}, {Zanin}, \& {Zapatero}}]{art:cygx1_tev}
---. 2007{\natexlab{b}}, \apjl, 665, L51, \dodoi{10.1086/521145}

\bibitem[{{Aleksi{\'c}} {et~al.}(2014){Aleksi{\'c}}, {Ansoldi}, {Antonelli}, {Antoranz}, {Babic}, {Bangale}, {Barrio}, {Becerra Gonz{\'a}lez}, {Bednarek}, {Bernardini}, {Biasuzzi}, {Biland}, {Blanch}, {Bonnefoy}, {Bonnoli}, {Borracci}, {Bretz}, {Carmona}, {Carosi}, {Colin}, {Colombo}, {Contreras}, {Cortina}, {Covino}, {Da Vela}, {Dazzi}, {De Angelis}, {De Caneva}, {De Lotto}, {de O{\~n}a Wilhelmi}, {Delgado Mendez}, {Dominis Prester}, {Dorner}, {Doro}, {Einecke}, {Eisenacher}, {Elsaesser}, {Fonseca}, {Font}, {Frantzen}, {Fruck}, {Galindo}, {Garc{\'\i}a L{\'o}pez}, {Garczarczyk}, {Garrido Terrats}, {Gaug}, {Godinovi{\'c}}, {Gonz{\'a}lez Mu{\~n}oz}, {Gozzini}, {Hadasch}, {Hanabata}, {Hayashida}, {Herrera}, {Hildebrand}, {Hose}, {Hrupec}, {Idec}, {Kadenius}, {Kellermann}, {Kodani}, {Konno}, {Krause}, {Kubo}, {Kushida}, {La Barbera}, {Lelas}, {Lewandowska}, {Lindfors}, {Lombardi}, {L{\'o}pez}, {L{\'o}pez-Coto}, {L{\'o}pez-Oramas}, {Lorenz}, {Lozano}, {Makariev}, {Mallot}, {Maneva}, {Mankuzhiyil}, {Mannheim},
  {Maraschi}, {Marcote}, {Mariotti}, {Mart{\'\i}nez}, {Mazin}, {Menzel}, {Miranda}, {Mirzoyan}, {Moralejo}, {Munar-Adrover}, {Nakajima}, {Niedzwiecki}, {Nilsson}, {Nishijima}, {Noda}, {Orito}, {Overkemping}, {Paiano}, {Palatiello}, {Paneque}, {Paoletti}, {Paredes}, {Paredes-Fortuny}, {Persic}, {Prada Moroni}, {Prandini}, {Puljak}, {Reinthal}, {Rhode}, {Rib{\'o}}, {Rico}, {Rodriguez Garcia}, {R{\"u}gamer}, {Saito}, {Saito}, {Satalecka}, {Scalzotto}, {Scapin}, {Schultz}, {Schweizer}, {Shore}, {Sillanp{\"a}{\"a}}, {Sitarek}, {Snidaric}, {Sobczynska}, {Spanier}, {Stamatescu}, {Stamerra}, {Steinbring}, {Storz}, {Strzys}, {Takalo}, {Takami}, {Tavecchio}, {Temnikov}, {Terzi{\'c}}, {Tescaro}, {Teshima}, {Thaele}, {Tibolla}, {Torres}, {Toyama}, {Treves}, {Uellenbeck}, {Vogler}, \& {Zanin}}]{3c58}
{Aleksi{\'c}}, J., {Ansoldi}, S., {Antonelli}, L.~A., {et~al.} 2014, \aap, 567, L8, \dodoi{10.1051/0004-6361/201424261}

\bibitem[{{Aliu} {et~al.}(2013){Aliu}, {Archambault}, {Arlen}, {Aune}, {Beilicke}, {Benbow}, {Bouvier}, {Buckley}, {Bugaev}, {Cesarini}, {Ciupik}, {Collins-Hughes}, {Connolly}, {Cui}, {Dickherber}, {Duke}, {Dumm}, {Dwarkadas}, {Errando}, {Falcone}, {Federici}, {Feng}, {Finley}, {Finnegan}, {Fortson}, {Furniss}, {Galante}, {Gall}, {Gillanders}, {Godambe}, {Gotthelf}, {Griffin}, {Grube}, {Gyuk}, {Hanna}, {Holder}, {Hughes}, {Humensky}, {Kaaret}, {Kargaltsev}, {Karlsson}, {Khassen}, {Kieda}, {Krawczynski}, {Krennrich}, {Lang}, {Lee}, {Madhavan}, {Maier}, {Majumdar}, {McArthur}, {McCann}, {Moriarty}, {Mukherjee}, {Nelson}, {O'Faol{\'a}in de Bhr{\'o}ithe}, {Ong}, {Orr}, {Otte}, {Park}, {Perkins}, {Pohl}, {Prokoph}, {Quinn}, {Ragan}, {Reyes}, {Reynolds}, {Roache}, {Roberts}, {Saxon}, {Schroedter}, {Sembroski}, {Slane}, {Smith}, {Staszak}, {Telezhinsky}, {Te{\v{s}}i{\'c}}, {Theiling}, {Thibadeau}, {Tsurusaki}, {Tyler}, {Varlotta}, {Vassiliev}, {Vincent}, {Vivier}, {Wakely}, {Weekes}, {Weinstein}, {Welsing},
  {Williams}, \& {Zitzer}}]{CTA1}
{Aliu}, E., {Archambault}, S., {Arlen}, T., {et~al.} 2013, \apj, 764, 38, \dodoi{10.1088/0004-637X/764/1/38}

\bibitem[{Ambrosone {et~al.}(2024)Ambrosone, Groth, Peretti, \& Ahlers}]{art:ambrosone_2024}
Ambrosone, A., Groth, K.~M., Peretti, E., \& Ahlers, M. 2024, Phys. Rev. D, 109, 043007, \dodoi{10.1103/PhysRevD.109.043007}

\bibitem[{{Amenomori} {et~al.}(2021){Amenomori}, {Bao}, {Bi}, {Chen}, {Chen}, {Chen}, {Chen}, {Chen}, {Cirennima}, {Danzengluobu}, {Fang}, {Fang}, {Feng}, {Feng}, {Feng}, {Gao}, {Gou}, {Guo}, {Guo}, {He}, {He}, {Hibino}, {Hotta}, {Hu}, {Hu}, {Huang}, {Jia}, {Jiang}, {Jin}, {Kasahara}, {Katayose}, {Kato}, {Kato}, {Kawata}, {Kihara}, {Ko}, {Kozai}, {Labaciren}, {Li}, {Li}, {Li}, {Lin}, {Liu}, {Liu}, {Liu}, {Liu}, {Liu}, {Lou}, {Lu}, {Meng}, {Munakata}, {Nakada}, {Nakamura}, {Nanjo}, {Nishizawa}, {Ohnishi}, {Ohura}, {Ozawa}, {Qian}, {Qu}, {Saito}, {Sakata}, {Sako}, {Shao}, {Shibata}, {Shiomi}, {Sugimoto}, {Takano}, {Takita}, {Tan}, {Tateyama}, {Torii}, {Tsuchiya}, {Udo}, {Wang}, {Wu}, {Xue}, {Yamamoto}, {Yang}, {Yokoe}, {Yuan}, {Zhai}, {Zhang}, {Zhang}, {Zhang}, {Zhang}, {Zhang}, {Zhang}, {Zhang}, {Zhao}, {Zhaxisangzhu}, \& {Tibet AS<SUB>{\ensuremath{\gamma}}</SUB> Collaboration}}]{amenomori}
{Amenomori}, M., {Bao}, Y.~W., {Bi}, X.~J., {et~al.} 2021, \prl, 126, 141101, \dodoi{10.1103/PhysRevLett.126.141101}

\bibitem[{{Archambault} {et~al.}(2017){Archambault}, {Archer}, {Benbow}, {Bird}, {Bourbeau}, {Buchovecky}, {Buckley}, {Bugaev}, {Cerruti}, {Connolly}, {Cui}, {Dwarkadas}, {Errando}, {Falcone}, {Feng}, {Finley}, {Fleischhack}, {Fortson}, {Furniss}, {Griffin}, {H{\"u}tten}, {Hanna}, {Holder}, {Johnson}, {Kaaret}, {Kar}, {Kelley-Hoskins}, {Kertzman}, {Kieda}, {Krause}, {Kumar}, {Lang}, {Maier}, {McArthur}, {McCann}, {Moriarty}, {Mukherjee}, {Nieto}, {O'Brien}, {Ong}, {Otte}, {Park}, {Pohl}, {Popkow}, {Pueschel}, {Quinn}, {Ragan}, {Reynolds}, {Richards}, {Roache}, {Sadeh}, {Santander}, {Sembroski}, {Shahinyan}, {Slane}, {Staszak}, {Telezhinsky}, {Trepanier}, {Tyler}, {Wakely}, {Weinstein}, {Weisgarber}, {Wilcox}, {Wilhelm}, {Williams}, \& {Zitzer}}]{Tycho}
{Archambault}, S., {Archer}, A., {Benbow}, W., {et~al.} 2017, Apj, 836, 23, \dodoi{10.3847/1538-4357/836/1/23}

\bibitem[{{Atwood} {et~al.}(2009){Atwood}, {Abdo}, {Ackermann}, {Althouse}, {Anderson}, {Axelsson}, {Baldini}, {Ballet}, {Band}, {Barbiellini}, {Bartelt}, {Bastieri}, {Baughman}, {Bechtol}, {B{\'e}d{\'e}r{\`e}de}, {Bellardi}, {Bellazzini}, {Berenji}, {Bignami}, {Bisello}, {Bissaldi}, {Blandford}, {Bloom}, {Bogart}, {Bonamente}, {Bonnell}, {Borgland}, {Bouvier}, {Bregeon}, {Brez}, {Brigida}, {Bruel}, {Burnett}, {Busetto}, {Caliandro}, {Cameron}, {Caraveo}, {Carius}, {Carlson}, {Casandjian}, {Cavazzuti}, {Ceccanti}, {Cecchi}, {Charles}, {Chekhtman}, {Cheung}, {Chiang}, {Chipaux}, {Cillis}, {Ciprini}, {Claus}, {Cohen-Tanugi}, {Condamoor}, {Conrad}, {Corbet}, {Corucci}, {Costamante}, {Cutini}, {Davis}, {Decotigny}, {DeKlotz}, {Dermer}, {de Angelis}, {Digel}, {do Couto e Silva}, {Drell}, {Dubois}, {Dumora}, {Edmonds}, {Fabiani}, {Farnier}, {Favuzzi}, {Flath}, {Fleury}, {Focke}, {Funk}, {Fusco}, {Gargano}, {Gasparrini}, {Gehrels}, {Gentit}, {Germani}, {Giebels}, {Giglietto}, {Giommi}, {Giordano}, {Glanzman},
  {Godfrey}, {Grenier}, {Grondin}, {Grove}, {Guillemot}, {Guiriec}, {Haller}, {Harding}, {Hart}, {Hays}, {Healey}, {Hirayama}, {Hjalmarsdotter}, {Horn}, {Hughes}, {J{\'o}hannesson}, {Johansson}, {Johnson}, {Johnson}, {Johnson}, {Johnson}, {Kamae}, {Katagiri}, {Kataoka}, {Kavelaars}, {Kawai}, {Kelly}, {Kerr}, {Klamra}, {Kn{\"o}dlseder}, {Kocian}, {Komin}, {Kuehn}, {Kuss}, {Landriu}, {Latronico}, {Lee}, {Lee}, {Lemoine-Goumard}, {Lionetto}, {Longo}, {Loparco}, {Lott}, {Lovellette}, {Lubrano}, {Madejski}, {Makeev}, {Marangelli}, {Massai}, {Mazziotta}, {McEnery}, {Menon}, {Meurer}, {Michelson}, {Minuti}, {Mirizzi}, {Mitthumsiri}, {Mizuno}, {Moiseev}, {Monte}, {Monzani}, {Moretti}, {Morselli}, {Moskalenko}, {Murgia}, {Nakamori}, {Nishino}, {Nolan}, {Norris}, {Nuss}, {Ohno}, {Ohsugi}, {Omodei}, {Orlando}, {Ormes}, {Paccagnella}, {Paneque}, {Panetta}, {Parent}, {Pearce}, {Pepe}, {Perazzo}, {Pesce-Rollins}, {Picozza}, {Pieri}, {Pinchera}, {Piron}, {Porter}, {Poupard}, {Rain{\`o}}, {Rando}, {Rapposelli}, {Razzano},
  {Reimer}, {Reimer}, {Reposeur}, {Reyes}, {Ritz}, {Rochester}, {Rodriguez}, {Romani}, {Roth}, {Russell}, {Ryde}, {Sabatini}, {Sadrozinski}, {Sanchez}, {Sander}, {Sapozhnikov}, {Parkinson}, {Scargle}, {Schalk}, {Scolieri}, {Sgr{\`o}}, {Share}, {Shaw}, {Shimokawabe}, {Shrader}, {Sierpowska-Bartosik}, {Siskind}, {Smith}, {Smith}, {Spandre}, {Spinelli}, {Starck}, {Stephens}, {Strickman}, {Strong}, {Suson}, {Tajima}, {Takahashi}, {Takahashi}, {Tanaka}, {Tenze}, {Tether}, {Thayer}, {Thayer}, {Thompson}, {Tibaldo}, {Tibolla}, {Torres}, {Tosti}, {Tramacere}, {Turri}, {Usher}, {Vilchez}, {Vitale}, {Wang}, {Watters}, {Winer}, {Wood}, {Ylinen}, \& {Ziegler}}]{atwood}
{Atwood}, W.~B., {Abdo}, A.~A., {Ackermann}, M., {et~al.} 2009, \apj, 697, 1071, \dodoi{10.1088/0004-637X/697/2/1071}

\bibitem[{{Bartoli} {et~al.}(2014){Bartoli}, {Bernardini}, {Bi}, {Branchini}, {Budano}, {Camarri}, {Cao}, {Cardarelli}, {Catalanotti}, {Chen}, {Chen}, {Creti}, {Cui}, {Dai}, {D'Amone}, {Danzengluobu}, {De Mitri}, {D'Ettorre Piazzoli}, {Di Girolamo}, {Di Sciascio}, {Feng}, {Feng}, {Feng}, {Gou}, {Guo}, {He}, {Hu}, {Hu}, {Iacovacci}, {Iuppa}, {Jia}, {Labaciren}, {Li}, {Liguori}, {Liu}, {Liu}, {Liu}, {Lu}, {Ma}, {Ma}, {Mancarella}, {Mari}, {Marsella}, {Martello}, {Mastroianni}, {Montini}, {Ning}, {Panareo}, {Perrone}, {Pistilli}, {Ruggieri}, {Salvini}, {Santonico}, {Shen}, {Sheng}, {Shi}, {Surdo}, {Tan}, {Vallania}, {Vernetto}, {Vigorito}, {Wang}, {Wu}, {Wu}, {Xue}, {Yang}, {Yang}, {Yao}, {Yuan}, {Zha}, {Zhang}, {Zhang}, {Zhang}, {Zhang}, {Zhao}, {Zhaxiciren}, {Zhaxisangzhu}, {Zhou}, {Zhu}, {Zhu}, {Zizzi}, \& {ARGO-YBJ Collaboration}}]{argo}
{Bartoli}, B., {Bernardini}, P., {Bi}, X.~J., {et~al.} 2014, \apj, 790, 152, \dodoi{10.1088/0004-637X/790/2/152}

\bibitem[{Bednarek(2003)}]{art:bednarek}
Bednarek, W. 2003, A\&A, 407, 1 , \dodoi{https://doi.org/10.1051/0004-6361:20030929}

\bibitem[{{Cao} {et~al.}(2021){Cao}, {Aharonian}, {An}, {Axikegu}, {Bai}, {Bao}, {Bastieri}, {Bi}, {Bi}, {Cai}, {Cai}, {Cao}, {Chang}, {Chang}, {Chang}, {Chen}, {Chen}, {Chen}, {Chen}, {Chen}, {Chen}, {Chen}, {Chen}, {Chen}, {Chen}, {Chen}, {Chen}, {Chen}, {Cheng}, {Cheng}, {Cui}, {Cui}, {Cui}, {Dai}, {Dai}, {Dai}, {Danzengluobu}, {della Volpe}, {D'Ettorre Piazzoli}, {Dong}, {Fan}, {Fan}, {Fan}, {Fang}, {Fang}, {Feng}, {Feng}, {Feng}, {Feng}, {Gao}, {Gao}, {Gao}, {Gao}, {Ge}, {Geng}, {Gong}, {Gou}, {Gu}, {Guo}, {Guo}, {Guo}, {Guo}, {Han}, {He}, {He}, {He}, {He}, {He}, {He}, {Heller}, {Hor}, {Hou}, {Hou}, {Hu}, {Hu}, {Hu}, {Hu}, {Huang}, {Huang}, {Huang}, {Huang}, {Huang}, {Ji}, {Ji}, {Jia}, {Jiang}, {Jiang}, {Jin}, {Kuleshov}, {Levochkin}, {Li}, {Li}, {Li}, {Li}, {Li}, {Li}, {Li}, {Li}, {Li}, {Li}, {Li}, {Li}, {Li}, {Li}, {Li}, {Li}, {Li}, {Liang}, {Liang}, {Lin}, {Liu}, {Liu}, {Liu}, {Liu}, {Liu}, {Liu}, {Liu}, {Liu}, {Liu}, {Liu}, {Liu}, {Liu}, {Liu}, {Liu}, {Liu}, {Long}, {Lu}, {Lv}, {Ma}, {Ma}, {Ma},
  {Mao}, {Masood}, {Mitthumsiri}, {Montaruli}, {Nan}, {Pang}, {Pattarakijwanich}, {Pei}, {Qi}, {Ruffolo}, {Rulev}, {S{\'a}iz}, {Shao}, {Shchegolev}, {Sheng}, {Shi}, {Song}, {Stenkin}, {Stepanov}, {Sun}, {Sun}, {Sun}, {Tam}, {Tang}, {Tian}, {Wang}, {Wang}, {Wang}, {Wang}, {Wang}, {Wang}, {Wang}, {Wang}, {Wang}, {Wang}, {Wang}, {Wang}, {Wang}, {Wang}, {Wang}, {Wang}, {Wang}, {Wang}, {Wang}, {Wang}, {Wang}, {Wei}, {Wei}, {Wei}, {Wen}, {Wu}, {Wu}, {Wu}, {Wu}, {Wu}, {Xi}, {Xia}, {Xia}, {Xiang}, {Xiao}, {Xiao}, {Xin}, {Xin}, {Xing}, {Xu}, {Xu}, {Xue}, {Yan}, {Yang}, {Yang}, {Yang}, {Yang}, {Yang}, {Yang}, {Yang}, {Yao}, {Yao}, {Ye}, {Yin}, {Yin}, {You}, {You}, {Yu}, {Yuan}, {Zeng}, {Zeng}, {Zeng}, {Zeng}, {Zha}, {Zhai}, {Zhang}, {Zhang}, {Zhang}, {Zhang}, {Zhang}, {Zhang}, {Zhang}, {Zhang}, {Zhang}, {Zhang}, {Zhang}, {Zhang}, {Zhang}, {Zhang}, {Zhang}, {Zhang}, {Zhang}, {Zhang}, {Zhang}, {Zhao}, {Zhao}, {Zhao}, {Zhao}, {Zhao}, {Zheng}, {Zheng}, {Zhou}, {Zhou}, {Zhou}, {Zhou}, {Zhou}, {Zhou}, {Zhu}, {Zhu}, {Zhu},
  {Zhu}, \& {Zuo}}]{lhaaso_2021}
{Cao}, Z., {Aharonian}, F.~A., {An}, Q., {et~al.} 2021, \nat, 594, 33, \dodoi{10.1038/s41586-021-03498-z}

\bibitem[{Cao {et~al.}(2023)Cao, Aharonian, An, Axikegu, Bai, Bao, Bastieri, Bi, Bi, Cai, Cao, Cao, Cao, Chang, Chang, Chen, Chen, Chen, Chen, Chen, Chen, Chen, Chen, Chen, Chen, Chen, Chen, Cheng, Cheng, Cui, Cui, Cui, Cui, Dai, Dai, Dai, Danzengluobu, della Volpe, Dong, Duan, Fan, Fan, Fang, Fang, Feng, Feng, Feng, Feng, Feng, Gabici, Gao, Gao, Gao, Gao, Gao, Gao, Ge, Geng, Giacinti, Gong, Gou, Gu, Guo, Guo, Guo, Guo, Han, He, He, He, He, He, Heller, Hor, Hou, Hou, Hou, Hu, Hu, Hu, Huang, Huang, Huang, Huang, Huang, Huang, Huang, Ji, Jia, Jia, Jiang, Jiang, Jiang, Jin, Kang, Ke, Kuleshov, Kurinov, Li, Li, Li, Li, Li, Li, Li, Li, Li, Li, Li, Li, Li, Li, Li, Li, Li, Li, Li, Liang, Liang, Lin, Liu, Liu, Liu, Liu, Liu, Liu, Liu, Liu, Liu, Liu, Liu, Liu, Liu, Liu, Lu, Luo, Lv, Ma, Ma, Ma, Mao, Min, Mitthumsiri, Mu, Nan, Neronov, Ou, Pang, Pattarakijwanich, Pei, Qi, Qi, Qiao, Qin, Ruffolo, S\'aiz, Semikoz, Shao, Shao, Shchegolev, Sheng, Shu, Song, Stenkin, Stepanov, Su, Sun, Sun, Sun, Tam, Tang, Tang, Tian, Wang,
  Wang, Wang, Wang, Wang, Wang, Wang, Wang, Wang, Wang, Wang, Wang, Wang, Wang, Wang, Wang, Wang, Wang, Wang, Wang, Wang, Wei, Wei, Wei, Wen, Wu, Wu, Wu, Wu, Wu, Xi, Xia, Xia, Xiang, Xiao, Xiao, Xin, Xin, Xing, Xiong, Xu, Xu, Xu, Xu, Xue, Yan, Yan, Yan, Yang, Yang, Yang, Yang, Yang, Yang, Yang, Yang, Yang, Yao, Yao, Ye, Yin, Yin, You, You, Yu, Yuan, Yue, Zeng, Zeng, Zeng, Zha, Zhang, Zhang, Zhang, Zhang, Zhang, Zhang, Zhang, Zhang, Zhang, Zhang, Zhang, Zhang, Zhang, Zhang, Zhang, Zhang, Zhang, Zhang, Zhao, Zhao, Zhao, Zhao, Zhao, Zheng, Zhou, Zhou, Zhou, Zhou, Zhou, Zhou, Zhou, Zhu, Zhu, Zhu, Zhu, \& Zuo}]{lhaaso_2023}
Cao, Z., Aharonian, F., An, Q., {et~al.} 2023, Phys. Rev. Lett., 131, 151001, \dodoi{10.1103/PhysRevLett.131.151001}

\bibitem[{{Cao} {et~al.}(2023){Cao}, {Aharonian}, {An}, {Axikegu}, {Bai}, {Bao}, {Bastieri}, {Bi}, {Bi}, {Cai}, {Cao}, {Cao}, {Cao}, {Chang}, {Chang}, {Chen}, {Chen}, {Chen}, {Chen}, {Chen}, {Chen}, {Chen}, {Chen}, {Chen}, {Chen}, {Chen}, {Chen}, {Cheng}, {Cheng}, {Cui}, {Cui}, {Cui}, {Cui}, {Dai}, {Dai}, {Dai}, {Danzengluobu}, {della Volpe}, {Dong}, {Duan}, {Fan}, {Fan}, {Fang}, {Fang}, {Feng}, {Feng}, {Feng}, {Feng}, {Feng}, {Gabici}, {Gao}, {Gao}, {Gao}, {Gao}, {Gao}, {Gao}, {Ge}, {Geng}, {Giacinti}, {Gong}, {Gou}, {Gu}, {Guo}, {Guo}, {Guo}, {Guo}, {Han}, {He}, {He}, {He}, {He}, {He}, {Heller}, {Hor}, {Hou}, {Hou}, {Hou}, {Hu}, {Hu}, {Hu}, {Huang}, {Huang}, {Huang}, {Huang}, {Huang}, {Huang}, {Huang}, {Ji}, {Jia}, {Jia}, {Jiang}, {Jiang}, {Jiang}, {Jin}, {Kang}, {Ke}, {Kuleshov}, {Kurinov}, {Li}, {Li}, {Li}, {Li}, {Li}, {Li}, {Li}, {Li}, {Li}, {Li}, {Li}, {Li}, {Li}, {Li}, {Li}, {Li}, {Li}, {Li}, {Li}, {Liang}, {Liang}, {Lin}, {Liu}, {Liu}, {Liu}, {Liu}, {Liu}, {Liu}, {Liu}, {Liu}, {Liu}, {Liu}, {Liu},
  {Liu}, {Liu}, {Liu}, {Lu}, {Luo}, {Lv}, {Ma}, {Ma}, {Ma}, {Mao}, {Min}, {Mitthumsiri}, {Mu}, {Nan}, {Neronov}, {Ou}, {Pang}, {Pattarakijwanich}, {Pei}, {Qi}, {Qi}, {Qiao}, {Qin}, {Ruffolo}, {S{\'a}iz}, {Semikoz}, {Shao}, {Shao}, {Shchegolev}, {Sheng}, {Shu}, {Song}, {Stenkin}, {Stepanov}, {Su}, {Sun}, {Sun}, {Sun}, {Tam}, {Tang}, {Tang}, {Tian}, {Wang}, {Wang}, {Wang}, {Wang}, {Wang}, {Wang}, {Wang}, {Wang}, {Wang}, {Wang}, {Wang}, {Wang}, {Wang}, {Wang}, {Wang}, {Wang}, {Wang}, {Wang}, {Wang}, {Wang}, {Wang}, {Wei}, {Wei}, {Wei}, {Wen}, {Wu}, {Wu}, {Wu}, {Wu}, {Wu}, {Xi}, {Xia}, {Xia}, {Xiang}, {Xiao}, {Xiao}, {Xin}, {Xin}, {Xing}, {Xiong}, {Xu}, {Xu}, {Xu}, {Xu}, {Xue}, {Yan}, {Yan}, {Yan}, {Yang}, {Yang}, {Yang}, {Yang}, {Yang}, {Yang}, {Yang}, {Yang}, {Yang}, {Yao}, {Yao}, {Ye}, {Yin}, {Yin}, {You}, {You}, {Yu}, {Yuan}, {Yue}, {Zeng}, {Zeng}, {Zeng}, {Zha}, {Zhang}, {Zhang}, {Zhang}, {Zhang}, {Zhang}, {Zhang}, {Zhang}, {Zhang}, {Zhang}, {Zhang}, {Zhang}, {Zhang}, {Zhang}, {Zhang}, {Zhang}, {Zhang},
  {Zhang}, {Zhang}, {Zhao}, {Zhao}, {Zhao}, {Zhao}, {Zhao}, {Zheng}, {Zhou}, {Zhou}, {Zhou}, {Zhou}, {Zhou}, {Zhou}, {Zhou}, {Zhu}, {Zhu}, {Zhu}, {Zhu}, \& {Zuo.}}]{lhaaso_cat}
{Cao}, Z., {Aharonian}, F., {An}, Q., {et~al.} 2023, arXiv e-prints, arXiv:2305.17030, \dodoi{10.48550/arXiv.2305.17030}

\bibitem[{{Cataldo} {et~al.}(2019){Cataldo}, {Pagliaroli}, {Vecchiotti}, \& {Villante}}]{cataldo}
{Cataldo}, M., {Pagliaroli}, G., {Vecchiotti}, V., \& {Villante}, F.~L. 2019, \jcap, 2019, 050, \dodoi{10.1088/1475-7516/2019/12/050}

\bibitem[{Di~Palma {et~al.}(2017)Di~Palma, Guetta, \& Amato}]{Palma_2017}
Di~Palma, I., Guetta, D., \& Amato, E. 2017, The Astrophysical Journal, 836, 159, \dodoi{10.3847/1538-4357/836/2/159}

\bibitem[{{Distefano} {et~al.}(2002){Distefano}, {Guetta}, {Waxman}, \& {Levinson}}]{Distefano2002}
{Distefano}, C., {Guetta}, D., {Waxman}, E., \& {Levinson}, A. 2002, \apj, 575, 378, \dodoi{10.1086/341144}

\bibitem[{Evoli {et~al.}(2008)Evoli, Gaggero, Grasso, \& Maccione}]{evoli}
Evoli, C., Gaggero, D., Grasso, D., \& Maccione, L. 2008, Journal of Cosmology and Astroparticle Physics, 2008, 018, \dodoi{10.1088/1475-7516/2008/10/018}

\bibitem[{Fang \& Murase(2018)}]{fang}
Fang, K., \& Murase, K. 2018, Nature Physics, 14, \dodoi{10.1038/s41567-017-0025-4}

\bibitem[{Gaisser \& Stanev(2006)}]{gaisser_2006}
Gaisser, T.~K., \& Stanev, T. 2006, Nuclear Physics A, 777, 98, \dodoi{https://doi.org/10.1016/j.nuclphysa.2005.01.024}

\bibitem[{Gonzalez-Garcia {et~al.}(2009)Gonzalez-Garcia, Halzen, \& Mohapatra}]{art:gonzalez}
Gonzalez-Garcia, M., Halzen, F., \& Mohapatra, S. 2009, Astroparticle Physics, 31, 437, \dodoi{https://doi.org/10.1016/j.astropartphys.2009.05.002.}

\bibitem[{Guetta \& Amato(2003)}]{art:guetta_amato}
Guetta, D., \& Amato, E. 2003, Astroparticle Physics, 19, 403, \dodoi{https://doi.org/10.1016/S0927-6505(02)00221-9}

\bibitem[{Guetta {et~al.}(2023)Guetta, Hillman, \& Valle}]{Guetta_2023}
Guetta, D., Hillman, Y., \& Valle, M.~D. 2023, Journal of Cosmology and Astroparticle Physics, 2023, 015, \dodoi{10.1088/1475-7516/2023/03/015}

\bibitem[{{H.E.S.S. Coll.} {et~al.}(2018){H.E.S.S. Coll.}, {Abdalla, H.}, {Abramowski, A.}, {Aharonian, F.}, {Ait Benkhali, F.}, {Angüner, E. O.}, {Arakawa, M.}, {Arrieta, M.}, {Aubert, P.}, {Backes, M.}, {Balzer, A.}, {Barnard, M.}, {Becherini, Y.}, {Becker Tjus, J.}, {Berge, D.}, {Bernhard, S.}, {Bernlöhr, K.}, {Blackwell, R.}, {Böttcher, M.}, {Boisson, C.}, {Bolmont, J.}, {Bonnefoy, S.}, {Bordas, P.}, {Bregeon, J.}, {Brun, F.}, {Brun, P.}, {Bryan, M.}, {Büchele, M.}, {Bulik, T.}, {Capasso, M.}, {Carrigan, S.}, {Caroff, S.}, {Carosi, A.}, {Casanova, S.}, {Cerruti, M.}, {Chakraborty, N.}, {Chaves, R. C. G.}, {Chen, A.}, {Chevalier, J.}, {Colafrancesco, S.}, {Condon, B.}, {Conrad, J.}, {Davids, I. D.}, {Decock, J.}, {Deil, C.}, {Devin, J.}, {deWilt, P.}, {Dirson, L.}, {Djannati-Ataï, A.}, {Domainko, W.}, {Donath, A.}, {Drury, L. O’C.}, {Dutson, K.}, {Dyks, J.}, {Edwards, T.}, {Egberts, K.}, {Eger, P.}, {Emery, G.}, {Ernenwein, J.-P.}, {Eschbach, S.}, {Farnier, C.}, {Fegan, S.}, {Fernandes, M. V.},
  {Fiasson, A.}, {Fontaine, G.}, {Förster, A.}, {Funk, S.}, {Füßling, M.}, {Gabici, S.}, {Gallant, Y. A.}, {Garrigoux, T.}, {Gast, H.}, {Gaté, F.}, {Giavitto, G.}, {Giebels, B.}, {Glawion, D.}, {Glicenstein, J. F.}, {Gottschall, D.}, {Grondin, M.-H.}, {Hahn, J.}, {Haupt, M.}, {Hawkes, J.}, {Heinzelmann, G.}, {Henri, G.}, {Hermann, G.}, {Hinton, J. A.}, {Hofmann, W.}, {Hoischen, C.}, {Holch, T. L.}, {Holler, M.}, {Horns, D.}, {Ivascenko, A.}, {Iwasaki, H.}, {Jacholkowska, A.}, {Jamrozy, M.}, {Jankowsky, D.}, {Jankowsky, F.}, {Jingo, M.}, {Jouvin, L.}, {Jung-Richardt, I.}, {Kastendieck, M. A.}, {Katarzyński, K.}, {Katsuragawa, M.}, {Katz, U.}, {Kerszberg, D.}, {Khangulyan, D.}, {Khélifi, B.}, {King, J.}, {Klepser, S.}, {Klochkov, D.}, {Kluźniak, W.}, {Komin, Nu.}, {Kosack, K.}, {Krakau, S.}, {Kraus, M.}, {Krüger, P. P.}, {Laffon, H.}, {Lamanna, G.}, {Lau, J.}, {Lees, J.-P.}, {Lefaucheur, J.}, {Lemière, A.}, {Lemoine-Goumard, M.}, {Lenain, J.-P.}, {Leser, E.}, {Lohse, T.}, {Lorentz, M.}, {Liu, R.},
  {López-Coto, R.}, {Lypova, I.}, {Marandon, V.}, {Malyshev, D.}, {Marcowith, A.}, {Mariaud, C.}, {Marx, R.}, {Maurin, G.}, {Maxted, N.}, {Mayer, M.}, {Meintjes, P.J.}, {Meyer, M.}, {Mitchell, A. M. W.}, {Moderski, R.}, {Mohamed, M.}, {Mohrmann, L.}, {Morå, K.}, {Moulin, E.}, {Murach, T.}, {Nakashima, S.}, {de Naurois, M.}, {Ndiyavala, H.}, {Niederwanger, F.}, {Niemiec, J.}, {Oakes, L.}, {O’Brien, P.}, {Odaka, H.}, {Ohm, S.}, {Ostrowski, M.}, {Oya, I.}, {Padovani, M.}, {Panter, M.}, {Parsons, R. D.}, {Paz Arribas, M.}, {Pekeur, N. W.}, {Pelletier, G.}, {Perennes, C.}, {Petrucci, P.-O.}, {Peyaud, B.}, {Piel, Q.}, {Pita, S.}, {Poireau, V.}, {Poon, H.}, {Prokhorov, D.}, {Prokoph, H.}, {Pühlhofer, G.}, {Punch, M.}, {Quirrenbach, A.}, {Raab, S.}, {Rauth, R.}, {Reimer, A.}, {Reimer, O.}, {Renaud, M.}, {de los Reyes, R.}, {Rieger, F.}, {Rinchiuso, L.}, {Romoli, C.}, {Rowell, G.}, {Rudak, B.}, {Rulten, C. B.}, {Safi-Harb, S.}, {Sahakian, V.}, {Saito, S.}, {Sanchez, D. A.}, {Santangelo, A.}, {Sasaki, M.},
  {Schandri, M.}, {Schlickeiser, R.}, {Schüssler, F.}, {Schulz, A.}, {Schwanke, U.}, {Schwemmer, S.}, {Seglar-Arroyo, M.}, {Settimo, M.}, {Seyffert, A. S.}, {Shafi, N.}, {Shilon, I.}, {Shiningayamwe, K.}, {Simoni, R.}, {Sol, H.}, {Spanier, F.}, {Spir-Jacob, M.}, {Stawarz, Ł.}, {Steenkamp, R.}, {Stegmann, C.}, {Steppa, C.}, {Sushch, I.}, {Takahashi, T.}, {Tavernet, J.-P.}, {Tavernier, T.}, {Taylor, A. M.}, {Terrier, R.}, {Tibaldo, L.}, {Tiziani, D.}, {Tluczykont, M.}, {Trichard, C.}, {Tsirou, M.}, {Tsuji, N.}, {Tuffs, R.}, {Uchiyama, Y.}, {van der Walt, D. J.}, {van Eldik, C.}, {van Rensburg, C.}, {van Soelen, B.}, {Vasileiadis, G.}, {Veh, J.}, {Venter, C.}, {Viana, A.}, {Vincent, P.}, {Vink, J.}, {Voisin, F.}, {Völk, H. J.}, {Vuillaume, T.}, {Wadiasingh, Z.}, {Wagner, S. J.}, {Wagner, P.}, {Wagner, R. M.}, {White, R.}, {Wierzcholska, A.}, {Willmann, P.}, {Wörnlein, A.}, {Wouters, D.}, {Yang, R.}, {Zaborov, D.}, {Zacharias, M.}, {Zanin, R.}, {Zdziarski, A. A.}, {Zech, A.}, {Zefi, F.}, {Ziegler, A.}, {Zorn,
  J.}, \& {Żywucka, N.}}]{art:hpgs2018}
{H.E.S.S. Coll.}, {Abdalla, H.}, {Abramowski, A.}, {et~al.} 2018, A\&A, 612, A1, \dodoi{10.1051/0004-6361/201732098}

\bibitem[{{H.E.S.S. Coll.} {et~al.}(2022){H.E.S.S. Coll.}, {Aharonian}, {Ait Benkhali}, {Ang{\"u}ner}, {Ashkar}, {Backes}, {Baghmanyan}, {Barbosa Martins}, {Batzofin}, {Becherini}, {Berge}, {Bernl{\"o}hr}, {Bi}, {B{\"o}ttcher}, {Boisson}, {Bolmont}, {de Bony de Lavergne}, {Breuhaus}, {Brose}, {Brun}, {Caroff}, {Casanova}, {Cerruti}, {Chand}, {Chen}, {Cotter}, {Damascene Mbarubucyeye}, {Djannati-Ata{\"\i}}, {Dmytriiev}, {Doroshenko}, {Duffy}, {Egberts}, {Ernenwein}, {Fegan}, {Feijen}, {Fiasson}, {Fichet de Clairfontaine}, {Fontaine}, {F{\"u}{\ss}ling}, {Funk}, {Gabici}, {Gallant}, {Ghafourizadeh}, {Giavitto}, {Giunti}, {Glawion}, {Glicenstein}, {Grondin}, {Hermann}, {Hinton}, {H{\"o}rbe}, {Hofmann}, {Hoischen}, {Holch}, {Holler}, {Horns}, {Huang}, {Jamrozy}, {Jankowsky}, {Jung-Richardt}, {Kasai}, {Katarzy{\'n}ski}, {Katz}, {Khangulyan}, {Kh{\'e}lifi}, {Klepser}, {Klu{\'z}niak}, {Komin}, {Konno}, {Kosack}, {Kostunin}, {Le Stum}, {Lemi{\`e}re}, {Lemoine-Goumard}, {Lenain}, {Leuschner}, {Lohse}, {Luashvili},
  {Lypova}, {Mackey}, {Malyshev}, {Malyshev}, {Marandon}, {Marchegiani}, {Marcowith}, {Mart{\'\i}-Devesa}, {Marx}, {Maurin}, {Meyer}, {Mitchell}, {Moderski}, {Mohrmann}, {Montanari}, {Moulin}, {Muller}, {Murach}, {Nakashima}, {de Naurois}, {Nayerhoda}, {Niemiec}, {Priyana Noel}, {O{\textquoteright}Brien}, {Ohm}, {Olivera-Nieto}, {de Ona Wilhelmi}, {Ostrowski}, {Panny}, {Panter}, {Parsons}, {Peron}, {Pita}, {Poireau}, {Prokhorov}, {Prokoph}, {P{\"u}hlhofer}, {Punch}, {Quirrenbach}, {Reichherzer}, {Reimer}, {Reimer}, {Renaud}, {Reville}, {Rieger}, {Rowell}, {Rudak}, {Rueda Ricarte}, {Ruiz-Velasco}, {Sahakian}, {Sailer}, {Salzmann}, {Sanchez}, {Santangelo}, {Sasaki}, {Sch{\"a}fer}, {Sch{\"u}ssler}, {Schutte}, {Schwanke}, {Senniappan}, {Shapopi}, {Simoni}, {Sinha}, {Sol}, {Specovius}, {Spencer}, {Stawarz}, {Steinmassl}, {Steppa}, {Takahashi}, {Tanaka}, {Taylor}, {Terrier}, {Thorpe-Morgan}, {Tsirou}, {Tsuji}, {Tuffs}, {Uchiyama}, {Unbehaun}, {van Eldik}, {van Soelen}, {Veh}, {Venter}, {Vink}, {Wagner}, {Werner},
  {White}, {Wierzcholska}, {Wong}, {Yusafzai}, {Zacharias}, {Zargaryan}, {Zdziarski}, {Zech}, {Zhu}, {Zouari}, \& {{\.Z}ywucka}}]{Hess2022}
{H.E.S.S. Coll.}, {Aharonian}, F., {Ait Benkhali}, F., {et~al.} 2022, Science, 376, 77, \dodoi{10.1126/science.abn0567}

\bibitem[{{Hunter} {et~al.}(1997){Hunter}, {Bertsch}, {Catelli}, {Dame}, {Digel}, {Dingus}, {Esposito}, {Fichtel}, {Hartman}, {Kanbach}, {Kniffen}, {Lin}, {Mayer-Hasselwander}, {Michelson}, {von Montigny}, {Mukherjee}, {Nolan}, {Schneid}, {Sreekumar}, {Thaddeus}, \& {Thompson}}]{hunter}
{Hunter}, S.~D., {Bertsch}, D.~L., {Catelli}, J.~R., {et~al.} 1997, \apj, 481, 205, \dodoi{10.1086/304012}

\bibitem[{Ke~Fang(2024)}]{art:fang_2024}
Ke~Fang, J. S. G. . F.~H. 2024, Nature Astronomy, 8, 241, \dodoi{https://doi.org/10.1038/s41550-023-02128-0}

\bibitem[{{Levinson} \& {Waxman}(2001)}]{levinson2001}
{Levinson}, A., \& {Waxman}, E. 2001, \prl, 87, 171101, \dodoi{10.1103/PhysRevLett.87.171101}

\bibitem[{Linden {et~al.}(2017)Linden, Auchettl, Bramante, Cholis, Fang, Hooper, Karwal, \& Li}]{linden}
Linden, T., Auchettl, K., Bramante, J., {et~al.} 2017, Phys. Rev. D, 96, 103016, \dodoi{10.1103/PhysRevD.96.103016}

\bibitem[{Lipari \& Vernetto(2018)}]{lipari}
Lipari, P., \& Vernetto, S. 2018, Phys. Rev. D, 98, 043003, \dodoi{10.1103/PhysRevD.98.043003}

\bibitem[{Malkov \& Drury(2001)}]{art:dsa}
Malkov, M.~A., \& Drury, L.~O. 2001, Rep. Prog. Phys., 64, 429, \dodoi{10.1088/0034-4885/64/4/201}

\bibitem[{{Mohrmann} {et~al.}(2022){Mohrmann}, {Ohm}, {Rauth}, {Specovius}, {van Eldik}, {Abdalla}, {Aharonian}, {Ait-Benkhali}, {Anguener}, {Arcaro}, {Armand}, {Armstrong}, {Ashkar}, {Backes}, {Baghmanyan}, {Barbosa Martins}, {Barnacka}, {Barnard}, {Batzofin}, {Becherini}, {Berge}, {Bernloehr}, {Bi}, {B{\"o}ttcher}, {Boisson}, {Bolmont}, {Bony (de)}, {Breuhaus}, {Brose}, {Brun}, {Bulik}, {Bylund}, {Cangemi}, {Caroff}, {Casanova}, {Catalano}, {Chambery}, {Chand}, {Chen}, {Cotter}, {Curlo}, {Dalgleish}, {Damascene Mbarubucyeye}, {Davids}, {Davies}, {Devin}, {Djannati-Ata{\"\i}}, {Dmytriiev}, {Donath}, {Doroshenko}, {Dreyer}, {Du Plessis}, {Duffy}, {Egberts}, {Einecke}, {Ernenwein}, {Fegan}, {Feijen}, {Fiasson}, {Fichet de Clairfontaine}, {Fontaine}, {Frans}, {Fuessling}, {Funk}, {Gabici}, {Gallant}, {Giavitto}, {Giunti}, {Glawion}, {Glicenstein}, {Grondin}, {Hattingh}, {Haupt}, {Hermann}, {Hinton}, {Hofmann}, {Hoischen}, {Holch}, {Holler}, {Horns}, {Huang}, {Huber}, {H{\"o}rbe}, {Jamrozy}, {Jankowsky},
  {Joshi}, {Jung}, {Kasai}, {Katarzynski}, {Katz}, {Khangulyan}, {Khelifi}, {Klepser}, {Kluzniak}, {Komin}, {Konno}, {Kosack}, {Kostunin}, {Kreter}, {Kukec Mezek}, {Kundu}, {Lamanna}, {Le Stum}, {Lemiere}, {Lemoine-Goumard}, {Lenain}, {Leuschner}, {Levy}, {Lohse}, {Luashvili}, {Lypova}, {Mackey}, {Majumdar}, {Malyshev}, {Malyshev}, {Marandon}, {Marchegiani}, {Marcowith}, {Mares}, {Marti'i-Devesa}, {Marx}, {Maurin}, {Meintjes}, {Meyer}, {Mitchell}, {Moderski}, {Montanari}, {Moore}, {Morris}, {Moulin}, {Muller}, {Murach}, {Nakashima}, {Naurois (de)}, {Nayerhoda}, {Davids}, {Niemiec}, {Noel}, {O'Brien}, {Oberholzer}, {Olivera-Nieto}, {Ona-Wilhelmi (de)}, {Ostrowski}, {Panny}, {Panter}, {Parsons}, {Peron}, {Pita}, {Poireau}, {Prokhorov}, {Prokoph}, {Puehlhofer}, {Punch}, {Quirrenbach}, {Reichherzer}, {Reimer}, {Reimer}, {Remy}, {Renaud}, {Reville}, {Rieger}, {Romoli}, {Rowell}, {Rudak}, {Rueda Ricarte}, {Ruiz Velasco}, {Sahakian}, {Sailer}, {Salzmann}, {Sanchez}, {Santangelo}, {Sasaki}, {Schaefer}, {Schutte},
  {Schwanke}, {Sch{\"u}ssler}, {Senniappan}, {Seyffert}, {Shapopi}, {Shiningayamwe}, {Simoni}, {Sinha}, {Sol}, {Spackman}, {Spencer}, {Spir-Jacob}, {Stawarz}, {Steenkamp}, {Stegmann}, {Steinmassl}, {Steppa}, {Sun}, {Takahashi}, {Tanaka}, {Tavernier}, {Taylor}, {Terrier}, {Thiersen}, {Thorpe-Morgan}, {Tluczykont}, {Tomankova}, {Tsirou}, {Tsuji}, {Tuffs}, {Uchiyama}, {van der Walt}, {van Rensburg}, {van Soelen}, {Vasileiadis}, {Veh}, {Venter}, {Vincent}, {Vink}, {V{\"o}lk}, {Wagner}, {Watson}, {Werner}, {White}, {Wierzcholska}, {Wong}, {Yassin}, {Yusafzai}, {Zacharias}, {Zanin}, {Zargaryan}, {Zdziarski}, {Zech}, {Zhu}, {Zmija}, {Zouari}, \& {{\.Z}ywucka}}]{Mohrmann2021}
{Mohrmann}, L., {Ohm}, S., {Rauth}, R., {et~al.} 2022, in 37th International Cosmic Ray Conference, 789, \dodoi{10.22323/1.395.0789}

\bibitem[{{Orellana, M.} {et~al.}(2007){Orellana, M.}, {Bordas, P.}, {Bosch-Ramon, V.}, {Romero, G. E.}, \& {Paredes, J. M.}}]{mq_orellana}
{Orellana, M.}, {Bordas, P.}, {Bosch-Ramon, V.}, {Romero, G. E.}, \& {Paredes, J. M.} 2007, A\&A, 476, 9, \dodoi{10.1051/0004-6361:20078495}

\bibitem[{Pagliaroli {et~al.}(2016)Pagliaroli, Evoli, \& Villante}]{pagliaroli}
Pagliaroli, G., Evoli, C., \& Villante, F. 2016, Journal of Cosmology and Astroparticle Physics, 2016, \dodoi{10.1088/1475-7516/2016/11/004}

\bibitem[{Povh {et~al.}(2004)Povh, Rith, Scholz, \& Zetsche}]{povh_2004}
Povh, B., Rith, K., Scholz, C., \& Zetsche, F. 2004, {Particles and nuclei: an introduction to the physical concepts; 4th ed.} (Berlin: Springer), \dodoi{10.1007/978-3-662-05432-1}

\bibitem[{Razzaque {et~al.}(2010)Razzaque, Jean, \& Mena}]{razzaque_2010}
Razzaque, S., Jean, P., \& Mena, O. 2010, Phys. Rev. D, 82, 123012, \dodoi{10.1103/PhysRevD.82.123012}

\bibitem[{Schwefer {et~al.}(2023)Schwefer, Mertsch, \& Wiebusch}]{schwefer}
Schwefer, G., Mertsch, P., \& Wiebusch, C. 2023, The Astrophysical Journal, 949, 16, \dodoi{10.3847/1538-4357/acc1e2}

\bibitem[{Shao {et~al.}(2023)Shao, Lin, \& Yang}]{Shao:2023aoi}
Shao, C., Lin, S., \& Yang, L. 2023, Phys. Rev. D, 108, L061305, \dodoi{10.1103/PhysRevD.108.L061305}

\bibitem[{{Shara}(1981)}]{Shara1981}
{Shara}, M.~M. 1981, \apj, 243, 926, \dodoi{10.1086/158657}

\bibitem[{Starrfield {et~al.}(2008)Starrfield, Iliadis, \& Hix}]{Starrfield2008}
Starrfield, S., Iliadis, C., \& Hix, W.~R. 2008, Thermonuclear processes, 2nd edn., Cambridge Astrophysics (Cambridge University Press), 77–101, \dodoi{10.1017/CBO9780511536168.006}

\bibitem[{{Starrfield} {et~al.}(1972){Starrfield}, {Truran}, {Sparks}, \& {Kutter}}]{Starrfield1972}
{Starrfield}, S., {Truran}, J.~W., {Sparks}, W.~M., \& {Kutter}, G.~S. 1972, \apj, 176, 169, \dodoi{10.1086/151619}

\bibitem[{{Steinberg} \& {Metzger}(2020)}]{Steinberg2020}
{Steinberg}, E., \& {Metzger}, B.~D. 2020, \mnras, 491, 4232, \dodoi{10.1093/mnras/stz3300}

\bibitem[{Vecchiotti {et~al.}(2023)Vecchiotti, Villante, \& Pagliaroli}]{art:vecchiotti_2023}
Vecchiotti, V., Villante, F.~L., \& Pagliaroli, G. 2023, The Astrophysical Journal Letters, 956, L44, \dodoi{10.3847/2041-8213/acff60}

\bibitem[{Vila \& Romero(2010)}]{mq_vila}
Vila, G.~S., \& Romero, G.~E. 2010, Monthly Notices of the Royal Astronomical Society, 403, 1457, \dodoi{10.1111/j.1365-2966.2010.16208.x}

\bibitem[{{Wakely} \& {Horan}(2008)}]{tev_cat}
{Wakely}, S.~P., \& {Horan}, D. 2008, in International Cosmic Ray Conference, Vol.~3, International Cosmic Ray Conference, 1341--1344

\bibitem[{Xin-Hua(2022)}]{art:lhaaso_exp}
Xin-Hua, M. e.~a. 2022, Chinese Phys. C, 46, L56, \dodoi{10.1088/1674-1137/ac3fa6}

\bibitem[{Yacobi {et~al.}(2014)Yacobi, Guetta, \& Behar}]{Yacobi_2014}
Yacobi, L., Guetta, D., \& Behar, E. 2014, The Astrophysical Journal, 793, 48, \dodoi{10.1088/0004-637X/793/1/48}

\bibitem[{Zegarelli {et~al.}(2022)Zegarelli, Celli, Capone, Gagliardini, Campion, \& Di~Palma}]{art:zegarelli}
Zegarelli, A., Celli, S., Capone, A., {et~al.} 2022, Phys. Rev. D, 105, 083023, \dodoi{10.1103/PhysRevD.105.083023}

\bibitem[{{Zhou} {et~al.}(2023){Zhou}, {Su}, {Yang}, {Chen}, {Sun}, {Jiang}, {Wang}, {Wang}, {Zhang}, {Xu}, {Yan}, {Yuan}, {Chen}, {Ao}, \& {Ma}}]{SNR_MC}
{Zhou}, X., {Su}, Y., {Yang}, J., {et~al.} 2023, \apjs, 268, 61, \dodoi{10.3847/1538-4365/acee7f}

\end{thebibliography}
\bibliographystyle{aasjournal}

\end{document}